\documentclass[format=acmlarge, screen]{acmart}
\usepackage{multirow}
\usepackage{pifont}

\AtBeginDocument{%
  \providecommand\BibTeX{{%
    \normalfont B\kern-0.5em{\scshape i\kern-0.25em b}\kern-0.8em\TeX}}}

\acmJournal{CSUR}
\acmVolume{0}
\acmNumber{0}
\acmMonth{0}

\usepackage{chngpage}
\usepackage{array,hhline}
\usepackage{flushend}
\usepackage{balance}
\usepackage{amsthm}

\usepackage{amsmath}
\usepackage{booktabs}
\usepackage{svg}
\usepackage{algorithmic}
\usepackage{makecell}
\usepackage{float}
\usepackage{stfloats}

\usepackage{amssymb}
\usepackage{array}
\usepackage{wasysym}

\usepackage{wrapfig}
\usepackage{stfloats}
\usepackage[outdir=./]{epstopdf}
\AtBeginDocument{%
  \providecommand\BibTeX{{%
    \normalfont B\kern-0.5em{\scshape i\kern-0.25em b}\kern-0.8em\TeX}}}


\setcopyright{none}
\copyrightyear{2022}
\acmYear{2022}
\acmDOI{10.1145/3547299}

\begin{document}

\title{Privacy Intelligence: A Survey on Image Privacy in Online Social Networks}
\author{Chi Liu}
\email{Chi.Liu@student.uts.edu.au}
\affiliation{
  \institution{University of Technology Sydney}
  \streetaddress{81 Broadway Ultimo}
  \city{Sydney}
  \state{NSW}
  \postcode{2007}
  \country{Australia}
}

\author{Tianqing Zhu}
\authornote{Tianqing Zhu is the corresponding author of this research.}
\email{Tianqing.zhu@uts.edu.au}
\affiliation{
  \institution{University of Technology Sydney}
  \streetaddress{81 Broadway Ultimo}
  \city{Sydney}
  \state{NSW}
  \postcode{2007}
  \country{Australia}
}

\author{Jun Zhang}
\email{junzhang@swin.edu.au}
\affiliation{%
  \institution{Swinburne University of Technology}
  \streetaddress{John St Hawthorn}
  \city{Sydney}
  \state{VIC}
  \postcode{3122}
  \country{Australia}
}

\author{Wanlei Zhou}
\email{wlzhou@cityu.edu.mo}
\affiliation{
  \institution{City University of Macau}
  \streetaddress{81 Av. Xian Xing Hai}
  \city{Macao}
  \country{China}
  }

\renewcommand{\shortauthors}{C.Liu et al.}

\begin{abstract}
Image sharing on online social networks (OSNs) has become an indispensable part of daily social activities, but it has also increased the risk of privacy invasion. An online image can reveal various types of sensitive information, prompting the public to rethink individual privacy needs in OSN image sharing critically. However, the interaction of images and OSN makes the privacy issues significantly complicated. The current real-world solutions for privacy management fail to provide adequate personalized, accurate and flexible privacy protection. Constructing a more intelligent environment for privacy-friendly OSN image sharing is urgent in the near future. Meanwhile, given the dynamics in both users' privacy needs and OSN context, a comprehensive understanding of OSN image privacy throughout the entire sharing process is preferable to any views from a single side, dimension or level. To fill this gap, we contribute a survey of "privacy intelligence" that targets modern privacy issues in dynamic OSN image sharing from a user-centric perspective. Specifically, we present the important properties and a taxonomy of OSN image privacy, along with a high-level privacy analysis framework based on the lifecycle of OSN image sharing. The framework consists of three stages with different principles of privacy by design. At each stage, we identify typical user behaviors in OSN image sharing and their associated privacy issues. Then a systematic review of representative intelligent solutions to those privacy issues is conducted, also in a stage-based manner. The analysis results in an intelligent "privacy firewall" for closed-loop privacy management. Challenges and future directions in this area are also discussed. 
\end{abstract}


\maketitle

\section{Introduction}
\label{section1}
\subsection{Background}
As an indispensable component of modern society, online social networks (OSNs) have significantly changed social communication and information exchange in the daily life of humans. Currently, sharing images on OSNs is extremely fashionable with its popularity is only growing. 
Photo capturing and posting can be completed with a few simple clicks anywhere and anytime, allowing users to instantly express themselves and interact with others. However, the convenience of sharing photos also raises considerable threats to users' privacy. Even a seemingly non-private OSN image can reveal significantly sensitive personal information. For instance, an online information forensics challenge launched in $2019$ \cite{NIXINTEL} showed that, by integrating multi-domain knowledge, 
a stranger could work out the photographer's sensitive information, such as when and where the picture was taken, from only one "ordinary" OSN photo of the user. 

In contrast to the simplicity of sharing images online, the preservation and management of OSN image privacy is remarkably challenging. There are three root causes, and the interplay of them makes the problem even harder.

\textbf{The intrinsic intractability of OSN image privacy.} The interaction of images with OSN introduces numerous new influence factors into the decision-making process of individual privacy. The privacy sensitivity of an online-shared image depends not only on the static image content but also on contextual dynamics from multiple dimensions, such as individuals' subjective decisions, the strength of social relationships, the nature of multi-party interactions in OSNs, and the spatial and temporal variations of an image \cite{Ahern:2007:OPP:1240624.1240683, Hoyle:PNP:2020}. These variables are intricately intertwined, making it impractical to address the problem merely from a single dimension, level, or side \cite{Hoyle:PNP:2020}. This intrinsic intractability of OSN image privacy significantly increases the difficulties with and cost of privacy management.  

\textbf{The natural limitations of human privacy consciousness.} OSN image privacy is subjective, closely associated with an individual's cognition and knowledge of privacy. Studies have shown that OSN users are ubiquitously unaware of how images can compromise privacy \cite{NYONI2018:PAU, MetaUnawareness}. Their common and seemingly innocuous behaviors in daily OSN image sharing can lead to unwitting violations of individual privacy for themselves and others, both directly and indirectly. And these infractions are beyond what the user could have imagined. Moreover, an individual's privacy preferences can be biased by, say, their education backgrounds, demographic characteristics, and social role \cite{acquisti2015privacy, Sunil:AEI, Hoyle:PNP:2020}, resulting in extraordinarily personalized privacy needs regarding OSN image sharing, especially when considering multiparty interactions \cite{Such:2017:PPC:3025453.3025668}.

\begin{wrapfigure}{Rh}{8cm}
\centering
\includegraphics[width=0.5\textwidth]{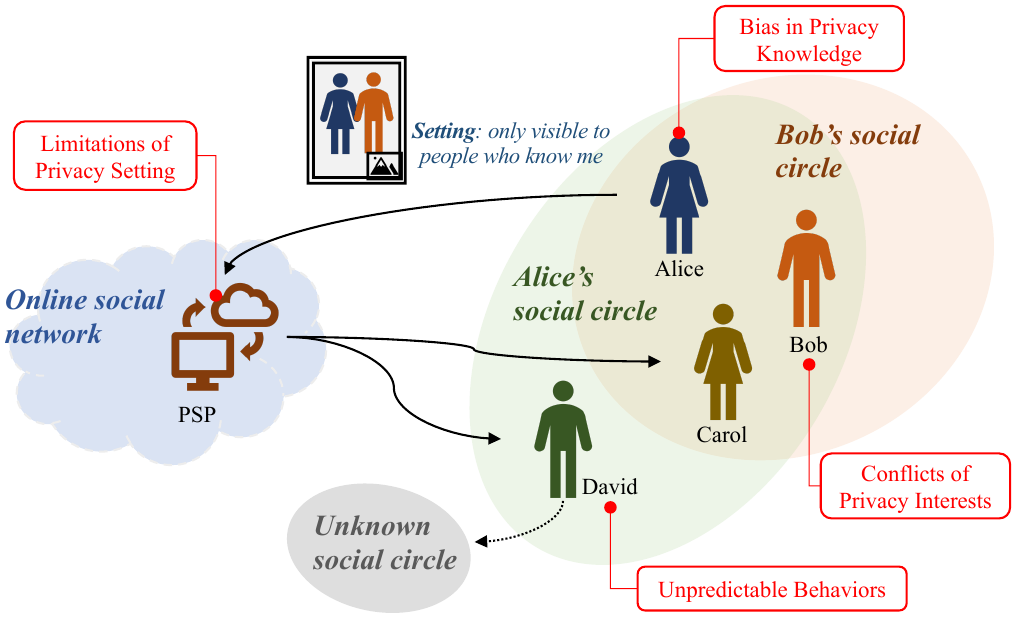}
\caption{A simplified OSN image sharing case. Several privacy risks (shown in the red boxes) still remain even if Alice enables the privacy setting for accessibility.}
\label{case2}
\end{wrapfigure}

\textbf{Machine learning exacerbates the tension between privacy and OSN image use.} Recent developments in machine learning have definitely made image sharing online more entertaining and enjoyable. But these techniques also pose great threats and intensify the difficulty of privacy preservation. The sensitive visual content of an image can be easily recognized or edited by a machine learning model \cite{Ding:2016:CSP:2885506.2845089}. Moreover, machine learning models can help reveal some implicit sensitive information such as occupation \cite{Chu:2016:POI:3012406.3009911}, health conditions \cite{Hossain2015:CAS} and even sexual orientation \cite{Wang:DNN:sex} from personal photos. Such machine learning-powered attacks are often cheap but hard to prevent in practice.

\textbf{\textit{Case study}: Does Alice really protect privacy as she had expected?}

\textit{A common OSN image sharing scenario (Figure \ref{case2}): The image owner Alice shares some photos about her and her partner Bobs' holiday trip to her own social circle. Supposing she already has certain privacy awareness and configures the privacy setting as "only visible to people who know me" (which is also very popular in practice).}


Despite Alice's privacy setting, we can easily see some privacy risks from different sides:

\textbf{Sender's side:} Alice may not know which information is truly sensitive and to what extent it can be revealed by today's novel attack methods. 

\textbf{Co-owner's side:} Bob's privacy may be violated if he does not want to disclose his relationship with Alice to people outside his own social circle, e.g., David. 

\textbf{Server's side:} No other fine-grained options are available by the current privacy setting system, albeit the social distances between Alice and each permitted visitor (e.g., Carol or David) may differ significantly.   

\textbf{Recipient's side:} The recipient (e.g., David) may be honest-but-curious and resend the photos to unknown people outside Alice's social circle, going against Alice's original privacy preference.

This is a simplified case, whereas, in reality, the situations are much more complicated, and the privacy risks are far greater than those listed above. In Sections \ref{section3} through \ref{section5}, we specify all privacy issues in OSN image sharing. 

\subsection{Motivation of privacy intelligence}
The above analysis and case study indicate considerable difficulties in dealing with the modern privacy issues associated with OSN image sharing. The current real-world managements, such as privacy laws and regulations \cite{Lynn:CUD, Victor:GDPR}, user agreements, or oversimplified privacy settings provided by OSN photo service providers (PSPs), are insufficient to eliminate the dilemma of OSN image privacy. Across the globe, this privacy crisis has spurred the research community to explore more effective solutions to satisfy modern privacy needs. The promising way requires further exploration and wider adoption of automated, semi-automated and human-computer-interactive techniques. We refer to this cluster of computer-aided privacy-enhancing technologies as \textbf{\emph{privacy intelligence}}. 

To date, various privacy intelligence solutions for OSN image sharing have been proposed. However, one concern remains that existing works mainly consider the problem from isolated views, solving each privacy issue individually. A survey to concatenate these fragments is critically needed to provide fundamental insights into this field and help build a privacy-friendly OSN image sharing environment. To this end, we review the literature published over the last decade surrounding this topic. Considering the user-centric aspiration of modern privacy management, we focus on the privacy needs associated with users' online image sharing operations. We begin with establishing a high-level privacy analysis framework based on the complete lifecycle of OSN image sharing. Typical user behaviors and their induced privacy issues are identified within the framework. Intelligent solutions targeting each issue are then discovered from the literature and analyzed comparatively. 

\subsection{Comparison of existing surveys}
There are several surveys partially related to this topic, including:
\begin{itemize}
    \item \textit{Surveys on OSN privacy.} Fire et al. \cite{Fire:OSNT} provided a thorough review of different security and privacy risks that threaten OSN users' well-being, along with an overview of existing countermeasures. Abawajy et al. \cite{Abawajy:PPS} presented a survey of privacy risks, attacks and privacy-preserving techniques in general for social network data publishing based on graph modelling methods. Alemany et al. \cite{alemany2022review} provided an in-depth analysis on privacy decision-making mechanisms in OSN.
    \item \textit{Surveys on multimedia privacy.} Padilla-L\'{o}pez et al. \cite{Jose:VPP} reviewed the protection techniques for visual data privacy, outlining useful design principles for privacy-aware intelligent monitoring systems. Ribaric et al. \cite{Ribaric:DPP} reviewed the de-identification mechanisms for non-biometric, physiological, behavioral, and soft-biometric identifiers in multimedia documents. Patsakis et al. \cite{Patsakis:PSM} outlined the significant security and privacy risks in exposing multimedia contents in OSN and discussed possible countermeasures. Zhang et al. \cite{zhang2022visual} surveyed typical attack and defense methods for visual privacy in deep learning systems.
\end{itemize}

Three key points differentiate our survey from these pioneering works. First, previous surveys mainly focus on static OSN privacy or local multimedia privacy independently, while we pay attention to the interplay of the two, considering privacy issues in the dynamic OSN image sharing context. Second, previous surveys were performed from a problem-driven perspective, i.e., identifying different threats and looking for specific countermeasures. In contrast, we try to establish a more global overview of this field from a user-centric perspective with a lifecycle-based privacy analysis framework. Last, previous surveys included many manpower-reliant solutions, while we purely focus on intelligent solutions, with an eye on interdisciplinary influences stemming from the recent progress of artificial intelligence. 

In brief, this article provides a comprehensive understanding of privacy in OSN image sharing with the following contributions: 1) the properties and taxonomy of OSN image privacy; 2) a novel privacy analysis framework based on the OSN image sharing lifecycle; 3) a representation of the potential privacy issues arising from typical sharing-associated user behaviors; 4) an in-depth review of the current advances in privacy intelligence solutions for OSN image privacy; 5) the identification of a set of essential design principles for privacy intelligence; 6) a discussion of the open challenges in OSN image privacy and future directions for addressing them.

    
    

The remainder of this paper is organized as follows. Section \ref{section2} presents an overview of OSN image privacy, discussing the properties and taxonomy of OSN image privacy and the structure of the privacy analysis framework. In Sections \ref{section3} through \ref{section5}, we identify modern privacy issues, summarize the corresponding intelligent solutions and discuss the common principles of privacy by design for each stage of the framework. Section \ref{section6} discusses the future lines for dealing with the open challenges detected. Section \ref{section7} offers a brief conclusion.

\section{OSN image privacy: an overview}\label{section2}

\subsection{Properties and taxonomy}
\label{definition}
There is no an exact definition of OSN image privacy in the current literature. As a starting point, we consider OSN image privacy with reference to the previous definitions on OSN privacy \cite{Nissenbaum:PCT, abraham2012:csn} and visual privacy \cite{Jose:VPP}: \textit{OSN image privacy is the \textbf{sensitivity},  \textbf{visibility} and \textbf{contextual integrity} of information exposed explicitly or implicitly from an image throughout the OSN propagation lifecycle.} With this definition, we highlight three key privacy properties:

\begin{itemize}
    \item \textbf{Sensitivity} defines what type of information associated with the image is sensitive in the sharing process.
    \item \textbf{Visibility} measures to what degree the sensitive information can be accessed by a specific recipient.
    \item \textbf{Contextual integrity} indicates the completeness of the information in the dynamic OSN image sharing context, which is governed by particular social norms.
\end{itemize}

\begin{wrapfigure}{Rh}{8cm}
\centering
\includegraphics[width=0.5\textwidth]{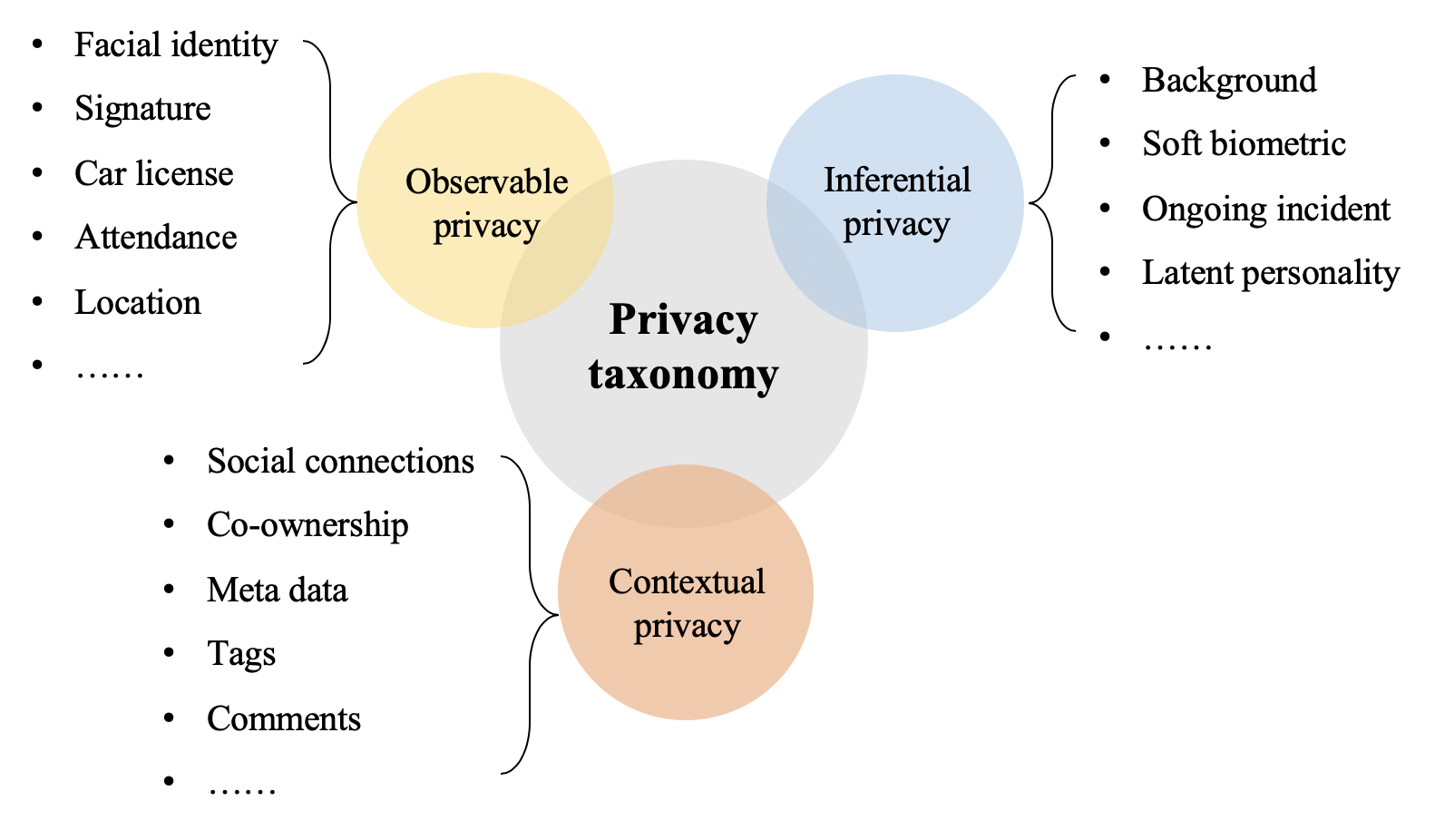}
\caption{A taxonomy of privacy in OSN image sharing}
\label{taxonomy}
\end{wrapfigure}

OSN image privacy can be categorized into three types shown in Figure \ref{taxonomy}.

\textbf{Observable privacy} refers to sensitive content that can be viewed directly in the image, such as faces, car licenses and location signs. This is the most common type of OSN image privacy, given the main purpose of online image sharing is to convey visual content to others for enjoyment. Observable privacy can be interpreted at different semantic levels, from lower ones such as pixel- and object-level, to higher ones like scene-level. It is the most vulnerable since the sensitive visual content is immediately exposed once images are accessed maliciously.

\textbf{Inferential privacy} refers to the sensitive information implied in the image content that can be inferred through reasoning or association. As an example, one might be able to extrapolate a specific event by analyzing the location cues in an image and/or what people are wearing. Soft biometric attributes, such as facial age and sexual orientation, are another typical type of inferential privacy. These underlying attributes are often hard for human viewers to perceive but can be deduced accurately by machines in the latent feature space \cite{Chu:2016:POI:3012406.3009911, Hossain2015:CAS, Wang:DNN:sex}.

\textbf{Contextual privacy} refers to the sensitive external information associated with an image in the dynamic OSN context. This type of information might be found in descriptive texts added by external actions during sharing, such as the metadata recorded by cameras or the auxiliary tags or captions provided by OSN users. It can also be found in the properties characterized by social interactions. For example, co-ownership of an image can be regarded as sensitive information since it can be exploited by attackers to reveal private social connections and possible image provenance and propagation paths.


\subsection{Privacy analysis framework}
\begin{figure*}[htb]
\centering
\includegraphics[width=\textwidth]{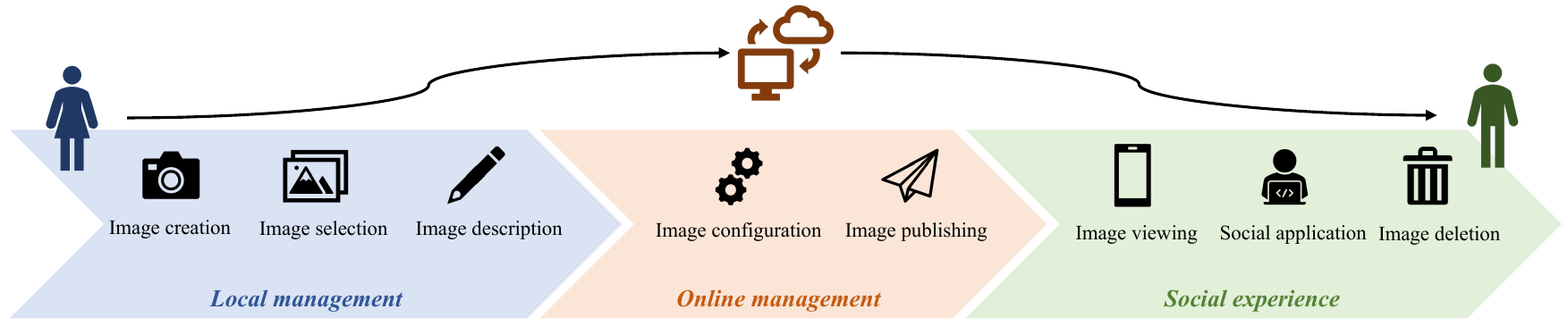}
\caption{The lifecycle-based privacy analysis framework, including three main stages: local management, online management and social experience. Each stage involves multiple sharing-specific user behaviors.}
\label{lifecycle}
\end{figure*}

We devised a privacy analysis framework to help identify diverse OSN image privacy issues and investigate privacy intelligence countermeasures. Considering the user-centric aspiration for modern privacy management, our focus is on the privacy issues induced by typical user behaviors in OSN image sharing. Hence, the framework covers all typical user behaviors in the entire lifecycle of OSN image sharing to form a closed loop for privacy analysis, as shown in Figure \ref{lifecycle}. There are three stages in this loop based on the progressive change in image controllability, including local management, online management and social experience. In each stage, the privacy solutions share common design principles of privacy intelligence, which will be discussed in the later sections. 

\subsubsection{User behaviors in local management} 
\ 
\newline
\indent \textbf{Image creation.} The image is captured via a camera. 

\textbf{Image selection.} The sender selects the image from a gallery for OSN sharing purposes.

\textbf{Image description.} Before publishing, the sender may attach particular information, such as descriptive tags, to the image to express a sentiment, foster social interaction or to help with image indexing.

\subsubsection{User behaviors in online management}
\ 
\newline
\indent \textbf{Image configuration.} Most current PSPs provide a configuration system for senders to manually set their privacy preferences regarding photo visibility. 

\textbf{Image publishing.} The sender publishes the image to make it accessible to the permitted recipients. 

\subsubsection{User behaviors in social experience}
\ 
\newline
\indent \textbf{Image viewing.}: The shared image is received and enjoyed by the authorized human recipients. 

\textbf{Social application.} The shared images are applied to specific online photo-based services. For example, OSN users might upload personal photos to a facial age evaluation application for fun. 

\textbf{Image deletion.} Users may choose to delete the image from OSNs after their sharing purposes have been met. Alternatively, some users may periodically or non-periodically check their image sharing records and delete some past photos to avoid long-term exposure online. 

\subsection{Technical goals of privacy intelligence} 
\label{ranking}
A user-friendly privacy intelligence solution in the context of OSN image sharing should never narrowly meet one single goal of privacy protection. Instead, we distil the following crucial technical goals of designing privacy intelligence through the literature review:

\indent \textbf{Privacy-utility trade-off (\textit{G1}-PU).} OSN users generally share an image with specific purposes, such as social interaction or moment recording. For whatever purposes, the shared image should maintain an essential level of utility after enhancing its privacy. 

\indent \textbf{Personalization (\textit{G2}-PE).} The solution is desired to be as personalized as possible to satisfy different OSN users' privacy needs.

\indent \textbf{Independence (\textit{G3}-IN).} A privacy intelligence solution is preferable not to rely on user data or third parties. Accessing user data such as historical records or user profiles is a privacy violation; the involvement of third parties leads to potential risks of information leakage. 

\indent \textbf{Automation (\textit{G4}-AU).} Automation means how "intelligent" the solution is. A more automated solution requires less human participation. 

\indent \textbf{Flexibility (\textit{G5}-FL).} Flexibility has two means; one is whether the solution can adaptively adjust to users' dynamics of users' privacy needs; the other one is whether the solution is compatible with present-day commercial OSN services or mobile devices. 

\indent \textbf{Communication-effectiveness (\textit{G6}-CE).} OSN image sharing is expected to be instant and low-latency. Thus, for those solutions involving multi-party communication, the time-effectiveness of communication is always a primary concern. 

We provide a qualitative comparison in terms of the above technical goals for each reviewed paper in Table \ref{tab:stage1}, \ref{tab:stage2} and \ref{tab:stage3}, with the following ranking system: \CIRCLE: The goal is a major concern with quantitative evaluations or has been fully satisfied in the paper. \LEFTcircle: The goal has been partially or weakly considered without quantitative evaluations in the paper. \Circle: The goal is not considered but worthy of careful concern in the problem formulated in the paper. \textbf{N/A}: The goal is not applicable to the problem formulated in the paper.


\section{Privacy analysis in local management}\label{section3}

\begin{figure*}[htb]
\centering
\includegraphics[width=\textwidth]{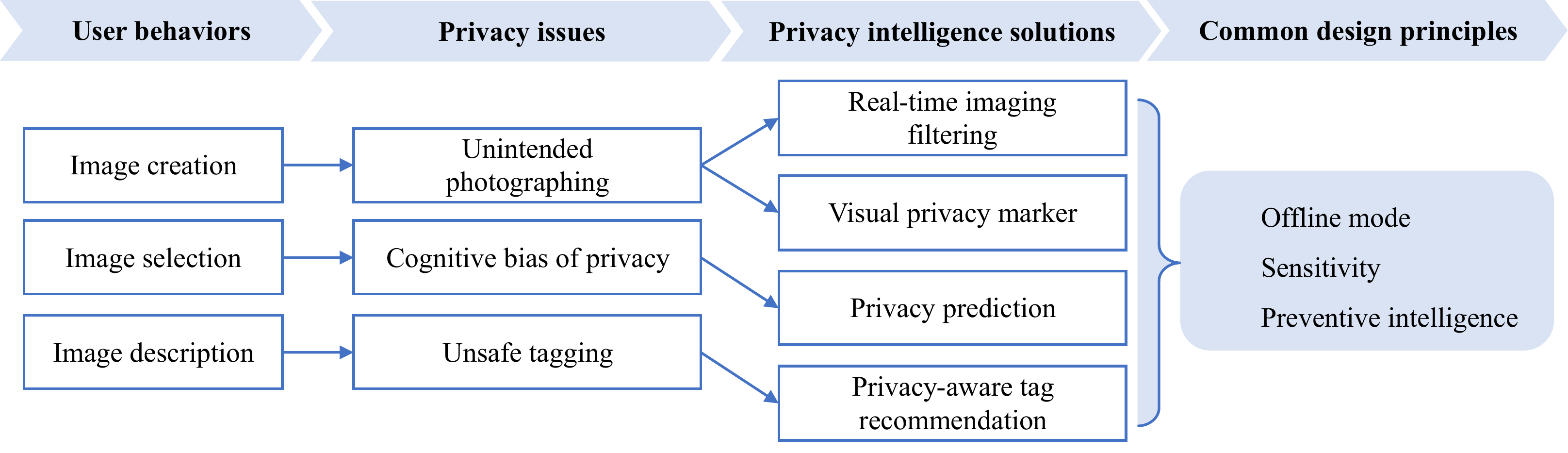}
\caption{Overview of the privacy analysis in the local management stage. Each privacy issue is linked to its user behavior cause and the corresponding intelligent solutions.}
\label{Stage1_fig}
\end{figure*}

This section provides a thorough privacy analysis in the local management stage, including privacy issues, intelligent solutions and the common design principles of privacy intelligence. Figure \ref{Stage1_fig} offers an overview of the analysis in this stage. 

\subsection{Privacy issues in local management}
\subsubsection{Issues associated with image capture}
\paragraph{\textbf{Unintended photographing.}}
When taking photos, it is sometimes inevitable that sensitive information of non-interested parties, such as bystanders' faces or military signs, are inadvertently captured. In most cases, users are unaware that the photographing behavior violates the privacy of non-interested parties, and even if they are, they can only delete or edit the photos manually according to their own understanding of privacy. 




\subsubsection{Issues associated with image selection}
\paragraph{\textbf{Cognitive bias of privacy.}} Previous studies have shown that users commonly lack a clear awareness of what images should be privacy-sensitive to themselves or related stakeholders. As a result, the selection of personal photos for online distribution often violates users' real privacy needs. This is also known as the "privacy paradox" \cite{kokolakis2017privacy}, which means users' actions have deviated from their true attitudes in pursuit of personal privacy.
 

\subsubsection{Issues associated with image description}
\paragraph{\textbf{Unsafe tagging.}} Image tagging has become a prevalent social functionality supported by many PSPs. However, indiscreet tagging behaviors might directly expose certain kinds of individual information, such as facial identity or location \cite{pesce2012privacy}. Automatic image tagging functions based on machine recognition algorithms may further exacerbate this risk \cite{chen2021automatic}. In addition, some PSPs adopt linkable tags as access control, which could lead to malicious access if the victim's face is deliberately tagged as another person \cite{tang2019faces}. 

\subsection{Privacy intelligence in local management}
\subsubsection{Solutions for unintended photographing}
\paragraph{\textbf{Real-time imaging filtering}}
Real-time imaging filtering is a solution that automatically identifies and removes sensitive content during the imaging process inside the camera. The implementations for real-time imaging filtering can be divided into hardware-based methods and software-based methods.

\textit{Hardware-based filtering.}\indent Chattopadhyay et al. \cite{Chattopadhyay} designed a digital signal processor for camera devices, which embeds an invertible cryptographic obscuration module to enhance privacy during photographing. The region of privacy interest of the captured image (defined as the person entering a static scene by the authors) was identified and encrypted using the block cipher algorithm Advanced Encryption Standard (AES) during image compression. Pittaluga et al. \cite{Pittaluga} proposed to filter the sensitive content in the pre-capture process via specific optical element design. A complementary optic layer in the camera sensor anonymizes the facial identity from the incident light field before sensor imaging. 

\textit{Software-based filtering.}\indent Aditya et al. \cite{Aditya:2016:IPP:2906388.2906412} developed I-Pic, a platform for individual policy-compliant content filtering in real-time photography. I-Pic maintains a secured signature based on facial features for each user to match his/her pre-defined privacy policies and photo presence. Then, once the user's presence in a non-related photo is detected, I-Pic will edit the photo according to his/her privacy policies. Another similar platform named Cardea is proposed by Shu et al. \cite{Shu:2018:CCV:3204949.3204973}. The difference between Cardea and I-Pic is that Cardea supports more fine-grained and context-aware privacy policies decided by four contextual elements: location, scene, others' presence, and hand gestures. 

Although both hardware and software-based methods can ensure privacy protection in real-time photography, they have some limitations. The hardware-based methods often lack flexibility, i.e., their designs always depend on pre-defined private content, such as human faces. In comparison, the software-based methods depend on individual privacy policies and are more personalized. However, the current systems such as I-Pic or Cardea require a third-party cloud for image management, which may be privacy-risky if the cloud provider is dishonest. 

\paragraph{\textbf{Visual privacy marker}} 
Visual privacy marker is another solution to prevent unintended photographing, where users use visual signs from the physical environment to express individual privacy needs when being captured. 

\textit{Natural visual marker.}\indent Schiff et al. \cite{Schiff2009} proposed assigning privacy-related meanings to physical objects (such as hats or vests). For example, people could wear specifically colored items to express their unwillingness to appear in a stranger's photos. The authors also developed a camera with an embedded visual tracker to automatically recognize these privacy hints when capturing an image.

\textit{Artificial visual marker.}\indent Pallas et al. \cite{Pallas:ONP} proposed using artificial visual markers instead of natural ones. They designed a set of four visual symbols representing four elementary privacy preferences. The idea is that photo subjects can wear these stickers in the form of stickers or badges that are easily recognizable. Bo et al. \cite{Bo:2014:PPC:2668332.2668339} designed a customized yet compatible QR code as a privacy marker. Compared with Pallas et al.'s design, the QR codes can be recognized more quickly and accurately by machines, and are more informative, allowing users to convey more personalized privacy needs.

Visual privacy markers allow users to express individual privacy needs directly and actively. This human-computer interactive design ensures great extendibility and applicability in real-world implementations. However, the current methods commonly require a build-in parsing module to translate the marker into structured policies, yet unified parsing rules are unavailable in the current industry. In addition, wearing a conspicuous artificial mark is not always practical for users.

\subsubsection{Solutions for cognitive bias of privacy}
\paragraph{\textbf{Privacy prediction}}
Human's cognitive bias regarding OSN image privacy often means manual privacy decisions are error-prone and time-consuming. A more efficient way is to automatically learn and predict the privacy pattern of an image with machine learning classifiers in a data-driven manner. The privacy pattern is normally formulated as a binary classification problem with a decision on "is privacy" or "is not privacy". An essential workflow can be determined: privacy-associated features are first determined and extracted, followed by a classifier trained to discover privacy patterns from the features.

\textit{Content-based prediction.}\indent Considering that image privacy is likely tied to visual content, Tran et al. \cite{Tran:2016:PFD:3015812.3016006} proposed extracting hierarchical features from visual content for privacy prediction. The object-level features and the general privacy features are extracted by two convolutional neural network (CNN)-based extractors respectively, and then concatenated for the final privacy decision. Han et al. \cite{han2022learning} proposed using multi-level and multi-scale features for privacy prediction. The multi-level features are extracted by different layers of a CNN and the multi-scale features by a stack of max-pooling layers. An self-attention-based \cite{vaswani2017attention} aggregation model was used for feature fusion and prediction. 

Some studies suggested learning the relatedness between semantic objects and privacy for privacy prediction. For example, Yu et al. \cite{Yu-iPrivacy} proposed an object-privacy alignment algorithm based on co-occurrence frequencies. The algorithm mines the object-privacy relatedness in a tree-based manner from massive social images. Then the tree-based features are combined with CNN features for detecting privacy-sensitive objects for privacy prediction. Yang et al. \cite{yang2020graph} built a knowledge graph from a large-scale image privacy dataset to represent the relevance between semantic objects and image privacy. A graph neural network is trained with the object-privacy graph for privacy prediction and explanation. Another graph-based prediction model developed by Yang et al. \cite{yang2022drag} employs a region-aware graph convolutional network to discover privacy-sensitive regions and model their correlation adaptively. A graph convolutional network combining the self-attention mechanism is adopted to model the dynamic interaction among the crucial regions identified from the spatially-correlated CNN feature maps. The local graph features are then concatenated with a global representation of the image to identify private images. 

\textit{Context-based prediction.}\indent OSN image privacy not only depends on the content but also on the associated OSN context (such as tags). In this vein, some researchers have explored the feasibility of using multi-modality contextual features for privacy prediction. Zerr et al. \cite{Zerr:2012:PIC:2348283.2348292} proposed fusing several hand-crafted visual features and textual features to train a privacy classifier. A similar feature fusion method was provided by Squicciarini et al. \cite{Squicciarini:2014:AIP:2631775.2631803}. The difference is that the authors additionally investigated the effectiveness of different feature combinations in privacy prediction. Their results revealed that the combination of scale-invariant features and tag features was the smallest best-performing set. Zhong et al. \cite{Zhong:2017:GPM:3172077.3172441} provides another insight using social group tendentiousness as a contextual feature with an assumption that the privacy decisions on the same image might vary in social groups with different privacy preferences. The expectation-maximization algorithm is used to estimate the likelihood for a user to be associated with each group according to users' historical privacy decisions and demographic information. Then the probability that any given image is private is a user-specific average of the privacy posteriors under each of the groups.

More recently, researchers have attempted to use multi-modality deep representative features instead of hand-crafted features for context-based prediction. Tonge et al. \cite{Tonge:2018:UDF:3184558.3191572} studied the usefulness of deep visual features and deep tag features for classifying image privacy. Deep visual features are represented as the multi-level CNN feature maps learned in an image classification task, while deep tag features are the CNN features associated with the top-$k$ objects in an object detection task. In a follow-up study \cite{Tonge:2019:DDM:3308558.3313691}, the authors proposed a dynamic ensemble learning algorithm which can rank the modality competence of different feature modalities including image objects, scenes and tags in privacy prediction. 

Learning-based privacy prediction can improve privacy decisions with data-informed intelligence that reduces human bias. There are two limitations in the current workflow. First, existing methods predicting image privacy from visual contents tend to link privacy with static semantic information, regardless of the dynamics of OSN image privacy. For example, most studies regard the human face as private, but one may feel that faces presented in a party photo are more sensitive than those taken in a public area. Thus, inter-object correlations and object-scene relationships should be carefully considered in feature selection. Second, most existing models formulate image privacy prediction as a "private v.s. public" classification problem. However, since users' privacy needs are subjective, it is impractical to define a distinct cut-off for privacy sensitivity. By contrast, adopting ranking scores or uncertainty probability \cite{liu2022label} are worthy of further investigation.

\subsubsection{Solutions for unsafe tagging}
\paragraph{\textbf{Privacy-aware tag recommendation}}
Privacy-aware tag recommendation is a set of machine-aided tag recommendation mechanisms aiming at automatically recommending high-quality privacy-aware tags for social images. Tonge et al. \cite{Tonge:2018:PTR:3209542.3209574} proposed a tag recommendation approach based on the correlations between tags and privacy patterns. The approach first identifies the top-$k$ neighboring images for the target image by both visual content similarity and tag similarity. Then, for each machine-detected tag of the target image, a ranking algorithm calculates the sum of similarities and the probability that the tag is private or not over the neighboring images, such that privacy-aware tags can be recommended empirically for the targeted image. Tang et al. \cite{tang2019faces} proposed a cooperative photo tagging system, with the aim of preventing malicious access through linkable tags. The system consists of two cascading stages: The first one is an initialization stage where new users' portrait samples are collected from their OSN profiles for identity matching and tagging; The second one is a cooperative tagging stage where the remaining unidentified participants are tagged cooperatively by the users identified in the first stage. This cooperative mode prevents malicious tagging behaviors from any single user side. 

\begin{table}[htbp]
\footnotesize
  \centering
  \caption{A summary of the solutions of privacy intelligence in the local management stage. OP: observable privacy; IP: inferential privacy; CP: contextual privacy. PU: privacy-utility trade-off; PE: personalization; IN: independence; AU: automation; FL: flexibility; CE: communication-effectiveness. The ranking system is explained in Section \ref{ranking}.}
    \begin{tabular}{l|l|l|l|l|l|l|l|l|l|l}
    \toprule
    \multicolumn{1}{p{8em}|}{Privacy intelligence } & \multicolumn{1}{p{2em}|}{Paper; Year} & \multicolumn{1}{p{6em}|}{Sub-class} & \multicolumn{1}{p{3.5em}|}{Target privacy} & \multicolumn{1}{p{7em}|}{Key technique} & \multicolumn{1}{p{1.5em}|}{\textit{G1}-PU} & \multicolumn{1}{p{1.5em}|}{\textit{G2}-PE} & \multicolumn{1}{p{1.5em}|}{\textit{G3}-IN} & \multicolumn{1}{p{1.5em}|}{\textit{G4}-AU} & \multicolumn{1}{p{1.5em}|}{\textit{G5}-FL} & \multicolumn{1}{p{1.5em}}{\textit{G6}-CE} \\
    \midrule
    \multicolumn{1}{l|}{\multirow{4}[8]{8em}{Real-time imaging filtering}} & \cite{Chattopadhyay}; 2007 & \makecell[l]{Hardware-based \\filtering} & OP    & \makecell[l]{Digital signal \\processor}  & \Circle   & \Circle   & \CIRCLE & \CIRCLE & \LEFTcircle   & \textbf{N/A} \\
\cmidrule{2-11}          & \cite{Pittaluga}; 2015 & \makecell[l]{Hardware-based \\filtering} & OP    & Optical imaging & \Circle   & \Circle   & \CIRCLE & \CIRCLE & \CIRCLE & \textbf{N/A} \\
\cmidrule{2-11}          & \cite{Aditya:2016:IPP:2906388.2906412}; 2016 & \makecell[l]{Software-based \\filtering} & OP    & \makecell[l]{Wireless \\communication}  & \Circle   & \LEFTcircle   & \Circle   & \LEFTcircle   & \CIRCLE & \CIRCLE \\
\cmidrule{2-11}          & \cite{Shu:2018:CCV:3204949.3204973}; 2018 & \makecell[l]{Software-based \\filtering} & OP    & \makecell[l]{Wireless \\communication}  & \Circle   & \CIRCLE & \Circle   & \LEFTcircle   & \CIRCLE & \CIRCLE \\
    \midrule
    \multicolumn{1}{l|}{\multirow{3}[6]{8em}{Visual privacy marker}} & \cite{Schiff2009}; 2009 & \makecell[l]{Natural visual \\marker} & OP    & \makecell[l]{Human-computer \\interaction }& \Circle   & \LEFTcircle   & \CIRCLE & \LEFTcircle   & \LEFTcircle   & \Circle \\
\cmidrule{2-11}          & \cite{Pallas:ONP}; 2014 & \makecell[l]{Artificial visual \\marker} & OP    & \makecell[l]{Human-computer \\interaction }& \Circle   & \LEFTcircle   & \CIRCLE & \LEFTcircle   & \LEFTcircle   & \Circle \\
\cmidrule{2-11}          & \cite{Bo:2014:PPC:2668332.2668339}; 2014 & \makecell[l]{Artificial visual \\marker} & OP    & \makecell[l]{Human-computer \\interaction }& \CIRCLE & \CIRCLE & \CIRCLE & \LEFTcircle   & \CIRCLE & \CIRCLE \\
    \midrule
    \multicolumn{1}{l|}{\multirow{10}[20]{8em}{Privacy prediction}} & \cite{Tran:2016:PFD:3015812.3016006}; 2016 & \makecell[l]{Content-based \\prediction} & OP; IP & \makecell[l]{Deep neural \\netwok} & \textbf{N/A}   & \Circle   & \CIRCLE & \CIRCLE & \Circle   & \textbf{N/A} \\
\cmidrule{2-11}          & \cite{han2022learning}; 2022 & \makecell[l]{Content-based \\prediction} & OP; IP & \makecell[l]{Deep neural \\netwok} & \textbf{N/A}   & \Circle   & \CIRCLE & \CIRCLE & \Circle   & \textbf{N/A} \\
\cmidrule{2-11}          & \cite{Yu-iPrivacy}; 2017 & \makecell[l]{Content-based \\prediction} & OP; IP & \makecell[l]{Deep multi-task \\learning}& \textbf{N/A}   & \Circle   & \CIRCLE & \CIRCLE & \Circle   & \textbf{N/A} \\
\cmidrule{2-11}          & \cite{yang2020graph}; 2020 & \makecell[l]{Content-based \\prediction} & OP; IP & \makecell[l]{Graph neural \\network} & \textbf{N/A}   & \Circle   & \CIRCLE & \CIRCLE & \Circle   & \textbf{N/A} \\
\cmidrule{2-11}          & \cite{yang2022drag}; 2022 & \makecell[l]{Content-based \\prediction} & OP; IP & \makecell[l]{Graph neural \\network} & \textbf{N/A}   & \Circle   & \CIRCLE & \CIRCLE & \Circle   & \textbf{N/A} \\
\cmidrule{2-11}          & \cite{Zerr:2012:PIC:2348283.2348292}; 2012 & \makecell[l]{Context-based \\prediction} & OP; IP & \makecell[l]{Feature \\engineering} & \textbf{N/A}   & \Circle   & \CIRCLE & \CIRCLE & \Circle   & \textbf{N/A} \\
\cmidrule{2-11}          & \cite{Squicciarini:2014:AIP:2631775.2631803}; 2014 & \makecell[l]{Context-based \\prediction} & OP; IP & \makecell[l]{Feature \\engineering} & \textbf{N/A}   & \Circle   & \CIRCLE & \CIRCLE & \Circle   & \textbf{N/A} \\
\cmidrule{2-11}          & \cite{Zhong:2017:GPM:3172077.3172441}; 2017 & \makecell[l]{Context-based \\prediction} & OP; IP & Machine learning & \textbf{N/A}   & \CIRCLE & \Circle   & \CIRCLE & \Circle   & \textbf{N/A} \\
\cmidrule{2-11}          & \cite{Tonge:2018:UDF:3184558.3191572}; 2018 & \makecell[l]{Context-based \\prediction} & OP; IP; CP & \makecell[l]{Deep neural \\netwok} & \textbf{N/A}   & \Circle   & \CIRCLE & \CIRCLE & \Circle   & \textbf{N/A} \\
\cmidrule{2-11}          & \cite{Tonge:2019:DDM:3308558.3313691}; 2019 & \makecell[l]{Context-based \\prediction} & OP; IP; CP & Ensemble learning & \textbf{N/A}   & \Circle   & \CIRCLE & \CIRCLE & \Circle   & \textbf{N/A} \\
    \midrule
    \multicolumn{1}{l|}{\multirow{2}[4]{8em}{Privacy-aware tag recommendation}} & \cite{Tonge:2018:PTR:3209542.3209574}; 2018 & -   & CP    & \makecell[l]{Deep neural \\netwok} & \Circle   & \LEFTcircle   & \Circle   & \CIRCLE & \Circle   & \textbf{N/A} \\
\cmidrule{2-11}          & \cite{tang2019faces}; 2019 & -   & OP    & Rule design  & \CIRCLE & \LEFTcircle   & \Circle   & \LEFTcircle   & \Circle   & \Circle \\
    \bottomrule
    \end{tabular}%
  \label{tab:stage1}%
\end{table}%

\subsection{Design principles in local management}
Table \ref{tab:stage1} provides a breakdown of the reviewed solutions in this stage. By examining the methods applied in these solutions, we can identify several common design principles regarding OSN image privacy in this stage.

\begin{itemize}
	\item \textbf{Offline mode.} In the local management stage, the image is prepared by the sender in an offline mode. In most cases, the image owner is the original sender with full control right of the image, and the image is only accessible by the sender. Therefore, all the reviewed solutions in this stage can be implemented in an offline mode. It offers the chance to integrate these solutions within a local modular that can be embedded into end camera applications or the initial user interfaces of OSN services. In this way, some privacy issues can be resolved immediately before getting further complicated in the subsequent OSN interactions. 
	
	\item \textbf{Sensitivity.} Recall the sensitivity property of OSN image privacy in Section \ref{definition}. This should be a primary design consideration in this stage. Since images are held and controlled by the owners with few online interactions with others, most privacy issues in this stage are derived from the owners' unawareness or knowledge limitations about OSN image privacy. Hence, one crucial goal is to assist users in identifying which kind of information in the image should be privacy-sensitive facing the OSN sharing process and help filter the sensitive information automatically in the presence of users' lack of privacy consciousness.

	\item \textbf{Preventive intelligence.} As discussed above,  the principal target of privacy intelligence in this stage is to automatically detect privacy-sensitive information to assist user decisions on OSN image sharing. Meanwhile, since the image owner has complete control of the image, it is preferable for most solutions to be performed in a human-computer-interactive fashion instead of automatically processing images directly. The computer alerts users of the privacy risks and offers appropriate recommendations before photo sharing online, forestalling photos from undesirable information leaks. In this way, privacy intelligence in this stage can be referred to as \emph{preventive intelligence}.
\end{itemize}


\section{Privacy analysis in online management}\label{section4}
\begin{figure*}[htbp]
\centering
\includegraphics[width=\textwidth]{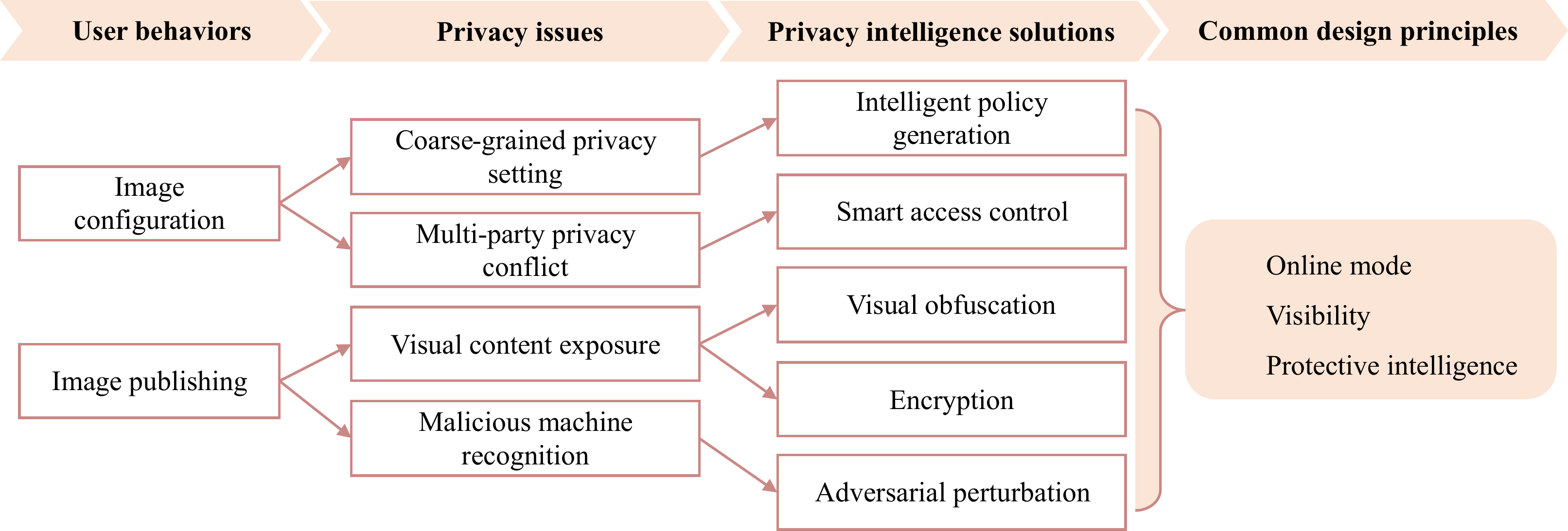}
\caption{Overview of the privacy analysis in the online management stage. Each privacy issue is linked to its user behavior cause and the corresponding intelligent solutions.}
\label{Stage2_fig}
\end{figure*}

This section focuses on privacy analysis in the online management stage.  Figure \ref{Stage2_fig} offers an overview of the analysis in this stage.


\subsection{Privacy issues in online management}
\subsubsection{Issues associated with image configuration}

\paragraph{\textbf{Coarse-grained privacy setting.}} The current privacy preference systems provided by most PSPs only support simple options, e.g., setting an image as publicly visible or private. Such coarse-grained settings cannot satisfy modern personalized individual privacy needs. Moreover, the configurations are heavily manpower-dependent, which is error-prone and tedious. 
 	
\paragraph{\textbf{Multi-party privacy conflicts.}} Normally, The shared images are closely related to multiple stakeholders. As users behave differently regarding how they disclose information \cite{Bart:DID}, they may have different opinions on what content is sensitive. Conflicts occur when the privacy settings of the sender override the privacy preferences of other stakeholders. The current OSNs cannot handle this issue effectively: the sender fully controls the configuration, whereas others are not granted any say in the matter. 

\subsubsection{Issues associated with image publishing}
\paragraph{\textbf{Visual content exposure.}} A common risk of online image publishing is the undesirable exposure of visual content. Users may get in trouble by disclosing sensitive visual content to unwanted viewers since it may cause a social impression suppression or economic loss. The unpredictability of the viewers' actions and morality further exacerbates this risk. 
 	
\paragraph{\textbf{Malicious machine recognition.}} Nowadays, the pervasive machine learning-based recognition systems bring considerable risks to individual privacy. Users' face photos can be easily collected online and applied to unknown face recognition systems. Meanwhile, the implicit information in images, such as health condition \cite{Hossain2015:CAS} and sex orientation \cite{Wang:DNN:sex}, can be accurately captured by machine learning models. These types of implicit information are associated with high-level image features, which is difficult to address with naive visual content processing. 


\subsection{Privacy intelligence in online management}
\subsubsection{Solutions for coarse-grained privacy setting}
\paragraph{\textbf{Intelligent policy generation}}
Intelligent policy generation is a cluster of computer-aided policy generation methods for mining fine-grained and personalized privacy policies according to environmental variations and user profiles. 

\textit{Rule-based policy.}\indent The descriptive information of OSN images such as tags and captions can be leveraged to design rules for privacy policy generation. Yeung et al. \cite{Yeung:PAC:2009} developed a policy recommendation mechanism using both tags and linked data provided by Semantic Web. The core idea is matching different groups in users' social circles to specific tags. For example, a photo tagged as "birthday" can only be accessed by a friend group. Klemperer et al. \cite{Klemperer:TYC:2012} investigated the usability of privacy rules based on tags created for organization, search, description, and communication. The rules are in the form "\textit{If tagged / not tagged with tag, then allow / deny}", combined with \textit{and} and \textit{or} as appropriate. The authors also found that when participants tagged photos with access control in mind, they were able to develop more coherent and accurate rules. 

\textit{Learning-based policy.}\indent Squicciarini et al. \cite{Squicciarini:PPI:2014} proposed a two-level learnable adaptive policy inference framework. The first-level component focuses on inferring individual privacy policies. It first learns to cluster a user's images using image content and metadata, then assigns initial rules to the images along with predicting the user's privacy tendencies. The second-level component adaptively adjusts the privacy policies over time by learning a multi-criteria inference network according to the user's historical social data and privacy attitudes. Yu et al. \cite{Yu:LCS:2017} suggested that content sensitivity and user trustworthiness are inseparable for determining privacy policy, and proposed a tree classifier-like policy generator exploiting the integration of the two types of information accordingly. The image content sensitivity is learned from both the deep representative features and the privacy-sensitive object features. User trustworthiness is represented by social group clustering based on the users' social behaviors. 

There are two challenges in the current policy generation models. First, since human relationships develop and the personal information exchange occurs in OSNs, mutual trustworthiness becomes more critical for the sharing activity. However, it is difficult to directly model real-world experience as numerical values in a virtual environment. One promising way is to employ automated tools to measure different metrics of trust in OSNs \cite{fogues2014bff, huynh2006integrated}. Second, the strength of the social connection between two OSN users may vary over time even when their friendship persists. Also, a user's understanding of privacy may change as their environment changes \cite{acquisti2015privacy}. A personalized policy generation method is therefore preferable to adapt to these changes immediately. However, many existing approaches ignore this dynamism because they are actualized by pre-defining a limited number of factors for policy learning.

\subsubsection{Solutions for multi-party privacy conflicts}
\paragraph{\textbf{Smart access control}}
Intelligent policy generation can be considered as a single-side access control mechanism that recommends policies to the image sender only. This solution is insufficient to address the problem of multi-party privacy conflicts since the sender-side policies are isolated from other stakeholders. A promising solution is smart access control, a set of mechanisms centralized on the PSP server to manage image accessibility considering all involved OSN users' interests.

\textit{Identity-based control.}\indent Some smart access control mechanisms leverage the personal identifiable information extracted from the shared image to decide accessibility. For example, Ilia et al. \cite{Ilia:FPP:2015} designed a facial identity-based access control model. The model exploits a three-dimensional relationship matrix involving users, photos, and faces in photos to manage multi-owner control policies. Each co-owner, i.e., the person depicted in the photo, sets a specific permissible viewer list as an entry of the relationship matrix. When an access request arrives, the model first identifies the co-owner's facial identity, then decides which face should be hidden according to the matrix. Similarly, Li et al. \cite{Li:HPP:2019} developed a framework that grants control rights to every co-owner. The difference is that Li et al.'s method provides an automated access control mechanism instead of setting policies photo-by-photo by users themselves. For a given photo, the mechanism identifies each co-owner and associates the photo with temporal, spatial, interpersonal, and attribute factors corresponding to the co-owner, to establish a scenario-level access control. Morris et al. \cite{morris2021you} proposed location-aware multi-party image access control mechanism allowing individual user to specify sensitive locations and timestamps for any photo in which their faces are identifiable. Each user pre-defined a privacy policy describing location range, location type, time and date interval and sensitiveness. Once a user is identified and the location of the photo is deemed sensitive, the user’s face will be replaced with a virtually generated human face according to the user’s privacy policy.

\textit{Social norm-based control.}\indent Another way for smart access control is employing social norms to regularize image propagation. Methods in this line usually formulates access control with the entire social graph $G \triangleq\langle V, E\rangle$, where $V$ is the set of users in this social network and $E$ is the set of edges connecting pairs of users with a specific relationship. Xu et al. \cite{Xu:TPP:2019} designed a trust-based access control model involving a collective incentive mechanism. This model relies on the maintainability of the mutual trust between OSN users. In each propagation round, the PSP selectively anonymizes the stakeholders according to their privacy loss and updates the trust value according to all stakeholders' feedback on whether their own faces have been anonymized correctly. Consequently, users taking others' privacy into account when sharing photos will gain more trust. Lin et al. \cite{Lin:REM:2020} proposed an access control mechanism by estimating the risk of image disclosure over unanticipated social graphs. Based on the image sharing historical data recorded by PSPs, the authors built a probability model that estimates the disclosure probabilities of different propagation channels, then aggregates the probability that other users will see the image shared by one user over various channels. The privacy policy of a given image is adjusted accordingly if the disclosure probability of this image is high.

\textit{Agent-based control.}\indent To respect co-owners' privacy, some works explored the multi-agent mechanism to collect agreements on sharing an image from the co-owners of the image. Kurtan et al. \cite{kurtan2021assisting} proposed an multi-agent approach where each user agent automatically predicts a policy for a new image according to the historical privacy policies of previous images. When in doubt, an agent analyzes the sharing behavior of other users in the same social network to recommend to its target user about what content should be private. Since each agent only accesses the privacy policies shared by users, this model is compatible with image-distributed environments. Motlagh et al. \cite{motlagh2021enabling} proposed an agent-based negotiation model available for both online and offline co-owners. The model consists of a coordination agent, a predictor agent, and a filtering algorithm. When an image is ready to be published, the coordinating agent associated with the sender collects opinions from online users and user agents operating on behalf of offline users. The predictor agent supports the user agents in opinion mining from previous opinions provided by affected users in similar contexts. Finally, the coordination agent uses the filtering algorithm to obscure all privacy-invasive information from the image based on the collected opinions. 

A persistent challenge with smart access control solutions is how to handle the privacy of all stakeholders to ensure everyone's privacy is respected as desired. Since a stakeholder's social relationship can be multiform (e.g., either interdependent or independent of the sender and the recipients), the complexity of the social graph rises significantly, making the multi-party access control more intractable. Meanwhile, the current access control mechanisms commonly underestimate the influence of social relationship \cite{relation1}. Most methods simplify social relationships into various groups, such as family, friends and colleagues, allocating a unified and static privacy policy to each group. However, it is more rational in practice that the relationship strength varies for each user pair and may change over time.


\subsubsection{Solutions for visual content exposure}
\begin{figure}[htbp]
\centering
\includegraphics[width = \textwidth]{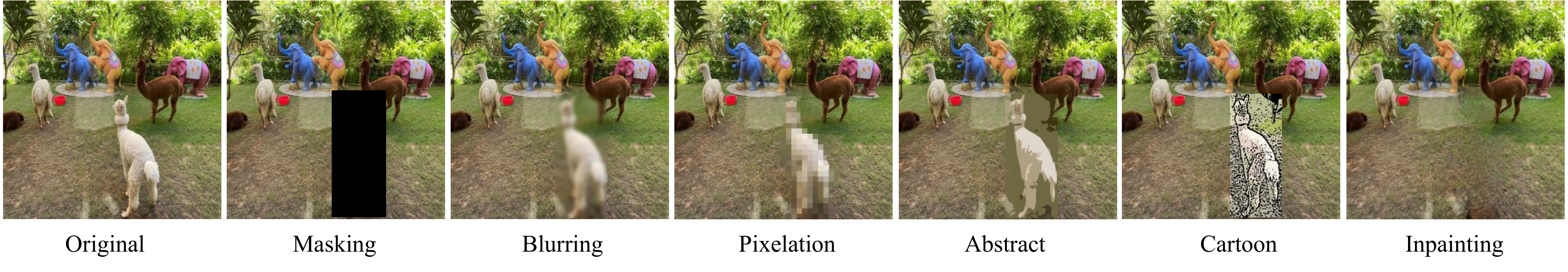}
\caption{Different visual obfuscation methods described in \cite{Li:ANF}.}
\label{visualob}
\end{figure}

\paragraph{\textbf{Visual obfuscation.}}
Visual obfuscation is widely used to prevent online images from undesirable exposure by hiding or removing sensitive regions with direct image modification. Figure \ref{visualob} shows examples of different visual obfuscation methods. The intuitive methods such as blurring, pixelation, cartooning and abstracting have been well summarized in previous surveys \cite{Jose:VPP, Patsakis:PSM}. Hence, our focus is on some of the emerging methods, including obfuscation with natural inpainting and obfuscation with measurable privacy. 

\textit{Natural obfuscation.}\indent One major concern of visual obfuscation is how to patching the modified region naturally to make the image realistic. Otherwise, the images' social utility will be compromised, and an adversary could easily perceive that the image has been edited. The images processed by intuitive methods such as blurring and pixelation cannot meet this need. Recently, some natural inpainting methods based on generative adversarial nets (GANs), a generative machine learning model that adversarially learns to map a target data distribution by contesting with a discriminative model \cite{Ian:GAN}, have emerged. These methods can seamlessly blend the scene surrounding the patched region to created authentic-looking images. 

For example, Uittenbogaard et al. \cite{Uittenbogaard:PPS} proposed an inpainting framework to remove privacy-sensitive pedestrians and vehicles in street-view imagery. The framework leverages the depth consistencies for detecting and removing objects. Next, a GAN employs multi-view information to inpaint the object-removed region. Sun et al. \cite{Sun:NEO} proposed a head inpainting approach for preserving facial identities. The challenge of head inpainting is that heads in photos usually appear with diverse motions and orientations. The authors address this problem by a GAN to generates sensible head and facial characteristics (e.g., facial landmarks) according to the image content (e.g., a body pose). Then another GAN synthesizes the non-existing heads aligned with the generated facial characteristics. In a subsequent study \cite{Sun:HMI} of the same team, the authors improved the head generator to support controllably different identities. The identity-related component of an original face is abstracted as a semantic parameter vector. Then the semantic parameters are modified and clustered into different identity groups, providing an explicit manipulation of identity. 
Kuang et al. \cite{kuang2021unnoticeable} proposed a seamless face replacement model based on the pix2pix model \cite{isola2017image} and U-Net \cite{ronneberger2015u}. By integrating the constraints over foreground (face) regions, background regions, and identity-related features, the model can synthesize faces that are naturally fused with the image background and significantly different from the original appearance. 

\textit{Privacy-measurable obfuscation.}\indent Another concern of conventional intuitive visual obfuscation methods is that they mostly manipulate the privacy-sensitive regions in a qualitative manner and fail to provide rigorous, quantitative and provable privacy guarantees. Recently, differential privacy (DP) \cite{zhu2020more}, a mathematical mechanism for measuring the privacy loss led by data publishing, has been studied for visual obfuscation. Fan et al. \cite{Fan:PIO} proposed an obfuscation approach based on metric privacy, a privacy measurement generalized from DP. The core idea is applying a sampling mechanism which satisfies metric privacy to the low-dimensional feature vectors transformed from the sensitive image region for privacy filtering. Li et al. \cite{Li:ANF} proposed a privacy-preserving attribute selection algorithm with privacy guarantees for facial image obfuscation. A set of facial attributes are identified by machine classifiers and then modified subject to $\epsilon$-DP constraint. A new face is reconstructed based on the anonymized attribute set with a GAN. Yu et al. \cite{yu2021gan} proposed to encode the privacy-sensitive objects into latent feature space using a GAN, and then introduce the Laplace noise into the latent features to ensure the DP-guaranteed de-identification.

Natural filling and rendering the obfuscated regions is crucial for retaining the images' social usability. However, high-quality filling and rendering may lead to another challenge of computational complexity, which is usually underestimated in the reviewed papers. Computational complexity is a particular concern for OSN services, given the necessity of time efficiency in real-time OSN image sharing. Most PSPs prefer traditional intuitive methods, such as blurring and pixilation, as these methods are easy to implement and have low latency. In comparison, sophisticated image reconstruction methods often require more computation in the feature space, which need to be further improved to adapt the OSN environment.

\paragraph{\textbf{Encryption.}}
Encryption is another automated solution widely adopted to avoid exposing undesirable visual content. The current methods can be divided into two bunches according to different technical goals. One is to maintain the recoverability of encryption in the presence of lossy image transformations. The other is to perform personalized and fine-grained encryption, which allows partial or hierarchical encryption according to recipients' access authority, rather than encrypting the entire image.

\textit{Recoverable encryption.}\indent OSNs noramlly apply various transformations to uploaded images for efficient storage and communication. However, these lossy operations can significantly affect the encryption/decryption performance. More resilient encryption/decryption methods are needed against these lossy transformations to ensure the lossless recovery. Tierney et al. \cite{Tierney:CPP:2013} proposed a photo encryption system tolerant to JPEG transformations. The authors defined $q,p$-Recoverability to guarantee the authorized recipient can decrypt the original image with a high probability $p$ under a minimum quality loss $q$. The system encrypts an image under a specific class of JPEG embedding protocols satisfying the $q,p$-Recoverability operated in the encrypted bit space. Sun et al. \cite{Sun:RPI:2018} considered the black-box problem where the lossy operations applied by OSNs are usually unknown to users and out of their control. Taking Facebook as a case, they estimated the parameters of four types of operations Facebook applies to the uploaded images through an offline training procedure, providing prior knowledge of developing a specific robust DCT-domain image encryption/decryption scheme.

Image steganography \cite{steganography}, a technology that hides cryptographic data in an image, offers another promising avenue for recoverable encryption. For example, Fu et al. \cite{Fu:ERD:2019} proposed a reversible data hiding scheme in encrypted images based on an adaptive encoding strategy. The original image content is encoded by block permutation and stream cipher. Then the most significant bits are identified for embedding additional data with reversed Huffman codes. With the encryption key and data hiding key, the recipient can extract hided data, decrypt and recover the image separately and efficiently. 

\textit{Personalized encryption.}\indent Another set of encryption methods focus on personalized content encryption, where the encryption strategies differ for different recipients according to the sender's privacy policy. For example, Ra et al. \cite{Ra:PTP:2013} proposed a secure image sharing system which separates an image into private and public parts by a signal component-based threshold. The system encrypted private component, leaving the remainder in a public and standards-compatible plaintext form. Different secure protocols were designed for the two parts: the authorized recipient can access both to recover the image, while others, such as the OSN provider, can only access the public part to perform server-side transformations to conserve bandwidth usage. A more fine-grained way is to hide the visual objects privacy-sensitive specific to recipients while keeping the remaining parts accessible. For instance, He et al. \cite{He:PTS:2016} proposed a partial encryption system for OSN images. The system leverages an automated detection and recommendation mechanism to determine the private part. It also allows users to customize the sensitive regions of their photos. The encryption is performed on DCT spectral coefficients, which is transparent to image transformations and can freely support most image processing libraries. 

Encryption can completely keep sensitive visual content from being disclosed to unauthenticated recipients. Most encryption methods assume an honest third party, commonly the OSN service provider, to maintain the encryption information such as cryptographic keys or communication protocols. However, this assumption is not always solid. Some OSN services may be interested in user photos for commercial purposes and therefore they may behave dishonestly. Further, these OSN services might be hacked. Hence, ways to perform encryption in extremely untrusted OSN environments is an open challenge.

\subsubsection{Solutions for malicious machine recognition}
\paragraph{\textbf{Adversarial perturbation.}}
\begin{wrapfigure}[]{R}{9cm}
\centering
\includegraphics[width = 0.55\textwidth]{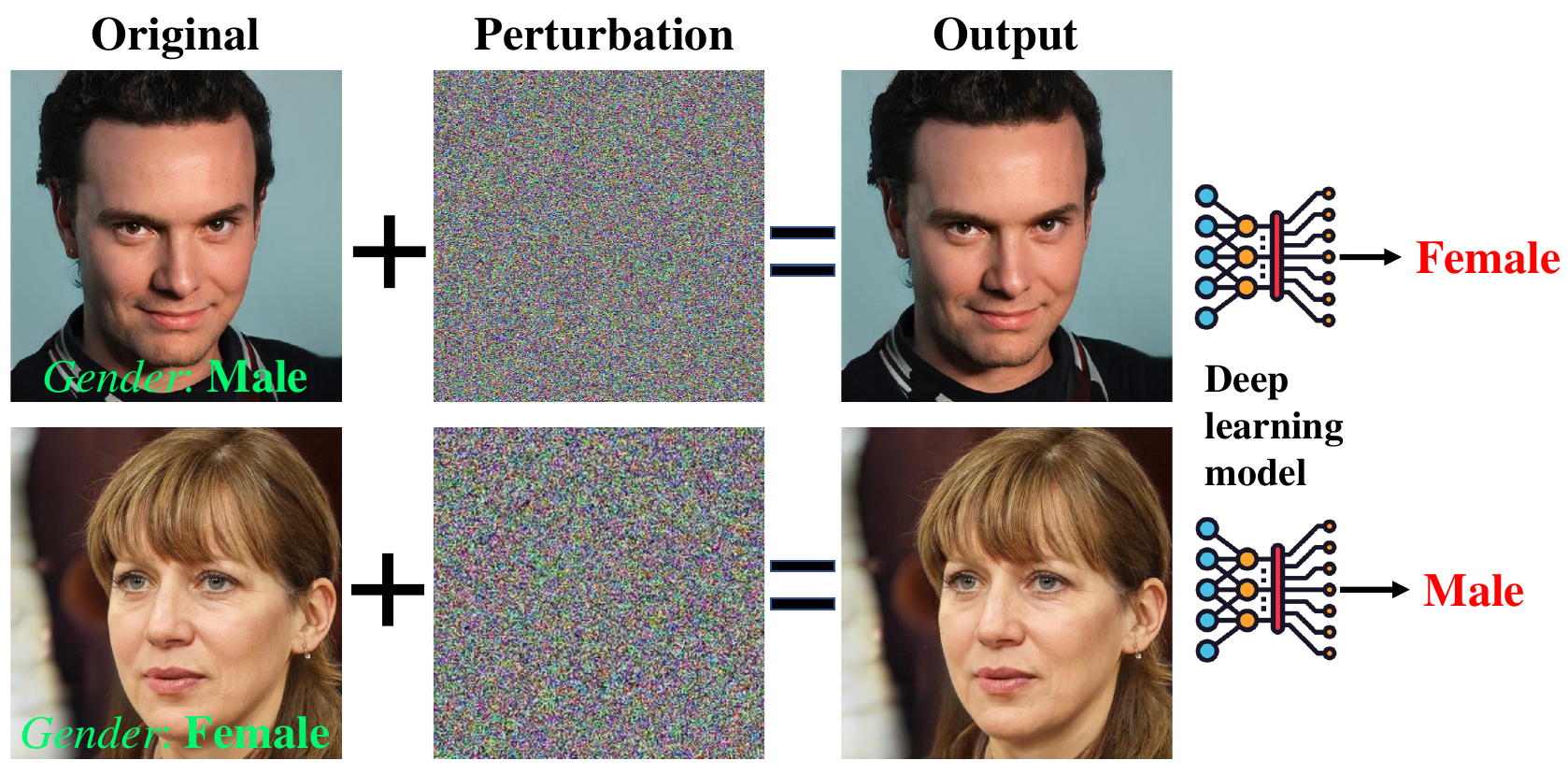}
\caption{An example of adversarial perturbation workflow described in \cite{mirjalili2017soft}. After adding perturbation noise to the original image, the gender recognition model is misled to make a wrong decision on the modified image. Both face images are synthesized by GAN.}
\label{ap_framework}
\end{wrapfigure}
Since machine recognition for images is often performed at the global-view level rather than at the object or region level, visual obfuscation of a region is impractical to defeat malicious machine recognition. Studies have shown that deep learning models can still make correct inferences on a partially obfuscated image \cite{DefeatingIO, Zhong:FED}. Therefore, solutions specific to machine adversaries are needed to prevent implicit attributes from disclosure. Recently, adversarial perturbation has shown efficacy in dealing with this problem. Adversarial perturbation is a technology that introduces imperceptible perturbations to an image, typically in the form of noise, in order to mislead deep learning recognition models \cite{moosavi2017universal}, as shown in Figure \ref{ap_framework}. 

\textit{Facial-level perturbation.}\indent Today's powerful machine learning-based facial recognition systems poses a real threat to OSN image privacy. To prevent user photos from unauthorized collections for training facial recognition models, Shan et al. \cite{shan2020fawkes} proposed an adversarial perturbation to add imperceptible pixel-level perturbation to user photos before sharing. The perturbations are computed via an image-specific optimization that maximizes the deviation of facial features for the target face recognition model. Then once these perturbed images are used as the training dataset for a facial recognition model, they can lead to a model consistently misidentifying the user's normal images.

More scholars focus on the inference phase of pre-trained unauthorized facial recognition models. A typical method is designing a perturbation objective function and optimizing it with the gradient sign method \cite{goodfellow2014explaining} to search for image-specific perturbations \cite{cherepanova2021lowkey, yang2021towards}. The optimization objective normally comprises two components: one is to make the target face recognition system predict the perturbed face as a different identity, which can be done by maximizing the identity feature distance between the target image and the candidate images of the same identity; the other one is minimizing the visual similarity between the perturbed face and the original face. Given that the real-world facial recognition models are often in black-box, an effective strategy is to perform the perturbation searching on an ensemble of surrogate white-box models \cite{cherepanova2021lowkey}. Another work from Shen et al. \cite{Shen:HPP:2019} further investigated how to make adversarial perturbation imperceptible. First, the authors conducted user studies to explore human sensitivity to feature-level visual changes on images, resulting in a sensitivity map that indicates the distribution of human visual sensitivity levels for a given image. With the guidance from the sensitivity map, the authors designed a sensitivity-aware adversarial perturbation model, which can precisely adjust the adversarial noise distribution to minimize visual distortion without compromising perturbation efficacy.  

\textit{Attribute-level perturbation.}\indent Images also contain sensitive auxiliary information related to implicit soft biometric attributes, such as age and gender, that machine learning models can recognize. Some works focused on perturbing biometric attributes in personal photos to resist unauthorized recognition while, optionally, preserving facial identity to maintain social usability. Mirjalili et al. \cite{mirjalili2017soft} proposed an adversarial perturbation algorithm to make an image evade gender recognition while retaining biometric identity. The algorithm iteratively perturbs face images using the gradient sign method \cite{goodfellow2014explaining}. In their follow-up study \cite{mirjalili2020privacynet}, the authors investigate how to conceal multiple biometric attributes simultaneously. A GAN-based perturbation model is designed to reconstruct images with selective attributes concealed automatically. The key idea is to embed the perturbation objectives and a visual consistency constraint in the GAN's training objective. Another GAN-based gender perturbation model is proposed by Tang et al. \cite{tang2022gender}. The aim is to obfuscate gender information while preserving the utility of other facial attributes. To this end, a selective transmission unit \cite{liu2019stgan} is employed in the latent feature space for disentangling gender attribute and other attributes, and a face matching loss is imposed to ensure the face recognition ability is well retained. Chhabra et al. \cite{Chhabra:AKF:2018} proposed a universal adversarial perturbation framework that can guarantee $k$-anonymity \cite{k-Anonymity} for selective facial attributes. The perturbation is learned with an objective function, ensuring that the $k$ facial attributes are anonymized as per the user's preferences. The framework also allows for a user control mechanism where users can select single or multiple attributes to be passed over. 

Applying adversarial perturbation to OSN images can effectively confuse deep learning models, and protect sensitive faces and facial attributes from unauthorized machine recognition. There are two challenges in this field. The first is transferability, which means the ability of an adversarial perturbation to remain effective even against a threat model other than the one originally targeted. The second challenge is universality, which indicates that a perturbation can fool a given model on any random images with high probability, also known as image-agnostic perturbation. Compared with the image-specific perturbation applied in most current methods, image-agnostic perturbation may be more feasible when confronting a large volume of images.

\subsection{Design principles in online management}
Table \ref{tab:stage2} provides a breakdown of the reviewed solutions in the online management stage. Similar to the first stage, several shared design principles regarding OSN image privacy can be identified in this stage.

\begin{table}[htbp]
\footnotesize
  \centering
  \caption{A summary of the solutions of privacy intelligence in the online management stage. OP: observable privacy; IP: inferential privacy; CP: contextual privacy. PU: privacy-utility trade-off; PE: personalization; IN: independence; AU: automation; FL: flexibility; CE: communication-effectiveness. The ranking system is explained in Section \ref{ranking}.}
    \begin{tabular}{l|l|l|l|l|l|l|l|l|l|l}
    \toprule
    \multicolumn{1}{p{5em}|}{\makecell[l]{Privacy\\ intelligence}} & \multicolumn{1}{p{2em}|}{Paper; Year} & \multicolumn{1}{p{5em}|}{Sub-class} & \multicolumn{1}{p{3.5em}|}{Target privacy} & \multicolumn{1}{p{8em}|}{Key technique} & \multicolumn{1}{p{1.5em}|}{\textit{G1}-PU} & \multicolumn{1}{p{1.5em}|}{\textit{G2}-PE} & \multicolumn{1}{p{1.5em}|}{\textit{G3}-IN} & \multicolumn{1}{p{1.5em}|}{\textit{G4}-AU} & \multicolumn{1}{p{1.5em}|}{\textit{G5}-FL} & \multicolumn{1}{p{1.5em}}{\textit{G6}-CE} \\
    \midrule
    \multicolumn{1}{l|}{\multirow{4}[8]{5em}{Intelligent policy generation}} & \cite{Yeung:PAC:2009}; 2009 & Rule-based policy & OP; CP & Rule design & \textbf{N/A}   & \LEFTcircle   & \CIRCLE & \CIRCLE & \CIRCLE & \textbf{N/A} \\
\cmidrule{2-11}          & \cite{Klemperer:TYC:2012}; 2012 & Rule-based policy & OP; CP & Rule design & \textbf{N/A}   & \LEFTcircle   & \CIRCLE & \LEFTcircle   & \Circle   & \textbf{N/A} \\
\cmidrule{2-11}          & \cite{Squicciarini:PPI:2014}; 2014 & Learning-based policy & OP; CP & Machine learning  & \textbf{N/A}   & \CIRCLE & \Circle   & \CIRCLE & \Circle   & \Circle \\
\cmidrule{2-11}          & \cite{Yu:LCS:2017}; 2017 & Learning-based policy & OP; CP & Deep learning & \Circle   & \CIRCLE & \Circle   & \CIRCLE & \Circle   & \Circle \\
    \midrule
    \multicolumn{1}{l|}{\multirow{7}[14]{5em}{Smart access control}} & \cite{Ilia:FPP:2015}; 2015 & Identity-based control & OP; CP & Rule mining & \Circle   & \LEFTcircle   & \CIRCLE & \LEFTcircle   & \CIRCLE & \CIRCLE \\
\cmidrule{2-11}          & \cite{Li:HPP:2019}; 2019 & Identity-based control & OP; IP; CP & Rule mining & \Circle   & \CIRCLE & \CIRCLE & \LEFTcircle   & \Circle   & \CIRCLE \\
\cmidrule{2-11}          & \cite{morris2021you}; 2021 & Identity-based control & OP; IP; CP & Rule mining & \CIRCLE & \LEFTcircle   & \CIRCLE & \CIRCLE  & \Circle   & \CIRCLE \\
\cmidrule{2-11}          & \cite{Xu:TPP:2019}; 2019 & Social norm-based control & OP; CP & Graph model & \Circle   & \LEFTcircle   & \Circle   & \LEFTcircle   & \Circle   & \Circle \\
\cmidrule{2-11}          & \cite{Lin:REM:2020}; 2020 & Social norm-based control & OP    & Graph model & \textbf{N/A}   & \LEFTcircle   & \Circle   & \CIRCLE & \LEFTcircle   & \Circle \\
\cmidrule{2-11}          & \cite{kurtan2021assisting}; 2021 & Agent-based control & OP    & Multi-agent model & \textbf{N/A}   & \CIRCLE & \Circle   & \LEFTcircle   & \Circle   & \Circle \\
\cmidrule{2-11}          & \cite{motlagh2021enabling}; 2021 & Agent-based control & OP; CP & Multi-agent model & \textbf{N/A}   & \CIRCLE & \Circle   & \LEFTcircle   & \Circle   & \Circle \\
    \midrule
    \multicolumn{1}{l|}{\multirow{7}[14]{5em}{Visual obfuscation}} & \cite{Uittenbogaard:PPS}; 2019  & Natural obfuscation & OP    & GAN & \CIRCLE & \Circle   & \CIRCLE & \CIRCLE & \Circle   & \textbf{N/A} \\
\cmidrule{2-11}          & \cite{Sun:NEO}; 2018 & Natural obfuscation & OP    & GAN & \CIRCLE & \Circle   & \CIRCLE & \CIRCLE & \Circle   & \textbf{N/A} \\
\cmidrule{2-11}          & \cite{Sun:HMI}; 2018 & Natural obfuscation & OP    & GAN & \CIRCLE & \Circle   & \CIRCLE & \CIRCLE & \Circle   & \textbf{N/A} \\
\cmidrule{2-11}          & \cite{kuang2021unnoticeable}; 2021 & Natural obfuscation & OP    & GAN & \CIRCLE & \Circle   & \CIRCLE & \CIRCLE & \Circle   & \textbf{N/A} \\
\cmidrule{2-11}          & \cite{Fan:PIO}; 2019 & \makecell[l]{Privacy-measurable \\obfuscation} & OP    & Metric privacy & \Circle   & \Circle   & \CIRCLE & \CIRCLE & \Circle   & \Circle \\
\cmidrule{2-11}          & \cite{Li:ANF}; 2019 & \makecell[l]{Privacy-measurable \\obfuscation} & OP    & Differential privacy & \CIRCLE & \CIRCLE & \CIRCLE & \CIRCLE & \Circle   & \Circle \\
\cmidrule{2-11}          & \cite{yu2021gan}; 2021 & \makecell[l]{Privacy-measurable \\obfuscation} & OP    & Differential privacy & \CIRCLE & \Circle   & \CIRCLE & \CIRCLE & \Circle   & \Circle \\
    \midrule
    \multicolumn{1}{l|}{\multirow{5}[10]{5em}{Encryption}} & \cite{Tierney:CPP:2013}; 2013 & Recoverable encryption & OP    & AES bit encryption & \CIRCLE & \Circle   & \CIRCLE & \CIRCLE & \Circle   & \CIRCLE \\
\cmidrule{2-11}          & \cite{Sun:RPI:2018}; 2018 & Recoverable encryption & OP    & DCT encryption & \CIRCLE & \Circle   & \CIRCLE & \CIRCLE & \Circle   & \CIRCLE \\
\cmidrule{2-11}          & \cite{Fu:ERD:2019}; 2019 & Recoverable encryption & OP    & Image steganography & \CIRCLE & \Circle   & \CIRCLE & \CIRCLE & \Circle   & \CIRCLE \\
\cmidrule{2-11}          & \cite{Ra:PTP:2013}; 2013 & Personalized encryption & OP    & DCT encryption & \CIRCLE & \CIRCLE & \Circle   & \LEFTcircle   & \LEFTcircle   & \CIRCLE \\
\cmidrule{2-11}          & \cite{He:PTS:2016}; 2016 & Personalized encryption & OP    & DCT encryption & \CIRCLE & \CIRCLE & \Circle   & \CIRCLE & \LEFTcircle   & \CIRCLE \\
    \midrule
    \multicolumn{1}{l|}{\multirow{8}[16]{5em}{Adversarial perturbation}} & \cite{shan2020fawkes}; 2020  & Facial-level perturbation & OP    & Adversarial attack & \CIRCLE & \Circle   & \CIRCLE & \CIRCLE & \Circle   & \textbf{N/A} \\
\cmidrule{2-11}          & \cite{cherepanova2021lowkey}; 2021 & Facial-level perturbation & OP    & Adversarial attack & \CIRCLE & \Circle   & \CIRCLE & \CIRCLE & \Circle   & \textbf{N/A} \\
\cmidrule{2-11}          & \cite{yang2021towards}; 2021 & Facial-level perturbation & OP    & Adversarial attack & \CIRCLE & \Circle   & \CIRCLE & \CIRCLE & \Circle   & \textbf{N/A} \\
\cmidrule{2-11}          & \cite{Shen:HPP:2019}; 2019 & Facial-level perturbation & OP    & Adversarial attack & \CIRCLE & \Circle   & \CIRCLE & \CIRCLE & \Circle   & \textbf{N/A} \\
\cmidrule{2-11}          & \cite{mirjalili2017soft}; 2017 & Attribute-level perturbation & IP    & Adversarial attack & \CIRCLE & \Circle   & \CIRCLE & \CIRCLE & \Circle   & \textbf{N/A} \\
\cmidrule{2-11}          & \cite{mirjalili2020privacynet}; 2020 & Attribute-level perturbation & IP    & GAN & \CIRCLE & \CIRCLE & \CIRCLE & \CIRCLE & \Circle   & \textbf{N/A} \\
\cmidrule{2-11}          & \cite{tang2022gender}; 2022 & Attribute-level perturbation & IP    & GAN & \CIRCLE & \CIRCLE & \CIRCLE & \CIRCLE & \Circle   & \textbf{N/A} \\
\cmidrule{2-11}          & \cite{Chhabra:AKF:2018}; 2018 & Attribute-level perturbation & IP    & Adversarial attack & \CIRCLE & \CIRCLE & \CIRCLE & \CIRCLE & \Circle   & \textbf{N/A} \\
    \bottomrule
    \end{tabular}%
  \label{tab:stage2}%
\end{table}%

\begin{itemize}
	\item \textbf{Online mode.} In the online management stage, the image is uploaded to the OSN server. In most cases, the image owner needs to allocate certain control rights to the server to configure and process the image. In this way, the OSN server is assumed to be willing to not only comply with the user's requirements but also to help actively manage image privacy. Ideally, all the reviewed solutions in this stage can be implemented and integrated into a PSP server as a cloud-based control engine. This centralized mechanism in the server can activate multiple protective solutions in an orderly and efficient manner, free from the user's perceptions or intervention. Meanwhile, an additional audit mechanism should be critically considered for OSN services whose honesty is in question.
	
	\item \textbf{Visibility.} The visibility property of OSN image privacy in Section \ref{definition} should be a primary design consideration in this stage. This property defines to what extent a specific recipient can access the sensitive information (identified in the local management stage). Hence, one fundamental goal of the solutions in this stage is to actualize feasible, efficient, and user-friendly access control in complex OSN contexts that can adjust the visibility of visual content or implicit attributes accordingly. 

	\item \textbf{Protective intelligence.} In contrast to the preventive intelligence of the local management stage, the principal target of privacy intelligence in this stage is to put in place tangible protections over the images to be released as per the user's need. The OSN server, which operates on the premise of having been empowered by the image owner, may conduct several privacy-enhancing operations directly on the image itself or in the image allocation and communication process. In this way, privacy intelligence in this stage can be referred to as \emph{ protective intelligence}.
\end{itemize}

\section{Privacy analysis in social experience}
\label{section5}
\begin{figure*}[htbp]
\centering
\includegraphics[width=\textwidth]{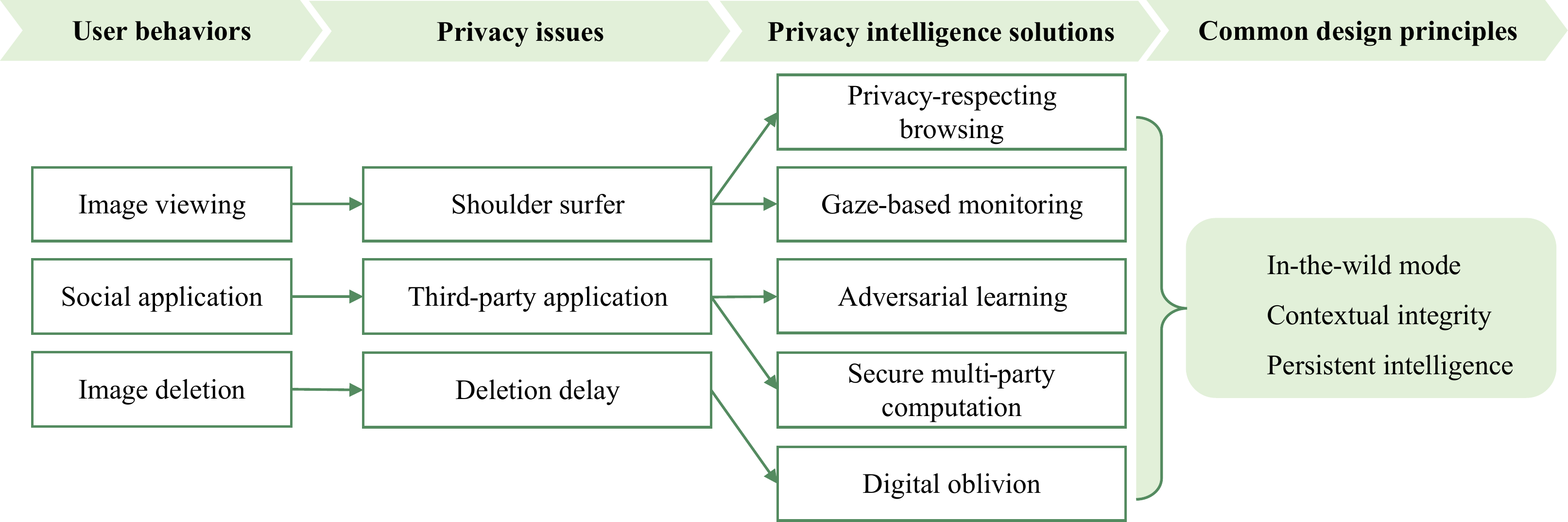}
\caption{Overview of the privacy analysis in the social experience stage. Each privacy issue is linked to its user behavior cause and the corresponding intelligent solutions.}
\label{Stage3_fig}
\end{figure*}

This section focuses on privacy analysis in the social experience stage. Figure \ref{Stage3_fig} offers an overview of the analysis in this stage. 

\subsection{Privacy issues in social experience}
\subsubsection{Issues associated with image viewing}

\paragraph{\textbf{Shoulder surfer.}} Shoulder surfer refers to the unauthorized viewers who visit private images by peeping at the victim's device screen over the victim's shoulder. This is a type of privacy intrusion from the physical world. Peoples' enthusiasm for viewing images on mobile devices further exacerbates this risk.

\subsubsection{Issues associated with social application}
\paragraph{\textbf{Third-party applications.}} One crucial purpose for modern OSN image sharing is to enjoy online third-party photo-based applications or services, such as facial emotion recognition or ancillary life loggers. These services usually employ automated recognition systems deployed in the cloud \cite{cloudApplication2, cloudApplication5}, where the image owner exercises little if any, control. An untrusted application may overstep its original access authority, performing excessive recognition beyond the user's request to steal personal information for commercial benefit \cite{Srivastava:2017:CUV:3131672.3131683}.


\subsubsection{Issues associated with image deletion}
\paragraph{\textbf{Deletion delay.}} The right to be forgotten is a critical modern privacy property required by the European General Data Protection Regulation (GDPR) \cite{Victor:GDPR}. Once the original sharing purposes are satisfied, the image should be removed from the Internet by specific strategies to avoid unknown distribution. However, the present OSN services easily suffer from deletion delay \cite{Liang:PCP}, which means an image exists much longer online than the user expects, even when the user's delete request has been executed \cite{FacebookDelete}.

\subsection{Intelligent solutions in social experience}

\subsubsection{Solutions for shoulder surfing}
\paragraph{\textbf{Privacy-respecting browsing}}
Humans are able to recognize images (especially face images) when they have seen the images before or know the face identity, even when the images are highly distorted \cite{Gregory:KPI}. Some studies exploit this human capability to develop privacy-respectful photo browsing approaches to prevent unknown shoulder surfers. Zezschwitz et al. \cite{Zezschwitz:YCW:2016} designed a new privacy-respectful digital reading pattern on smartphones to defend against unwanted observations. The proposed method distorts images in specific ways, rendering the visual content hard to recognize by an onlooker who does not know the photograph. By contrast, because the device owner knows the original photo and the details of how it was distorted, they have no problem recognizing them. Tajik et al. \cite{Tajik:BIP:2019} proposed an image transformation mechanism with a thumbnail-preserving encryption scheme. In this scheme, a ciphertext is defined as an image sharing the same thumbnail as the plaintext image but leaks nothing about the plaintext image beyond the thumbnail. With the proposed mechanism, users who know the original image can identify its encrypted version at the thumbnail level, while other people and even machine recognition systems cannot. By controlling the resolution of the thumbnail, users can obtain a good balance between online photo browsing and privacy.

The current privacy-respecting browsing methods assume that the viewer already knows the images they receive or the subjects depicted in the images. This assumption cannot apply to viewers seeing an image for the first time. In addition, the images are transformed into a low-quality format in the current methods. Although the viewers may recognize the transformed version, their viewing experience may be damaged by the low quality. Hence these methods have limited applicability: more suitable for online photo managing than viewing.

\paragraph{\textbf{Gaze-based monitoring}}
Gaze-based monitoring is an effective technique for privacy protection for public digital devices by tracking eye or motion movements, which has been exploited to protect people from shoulder surfing \cite{EyeGaze}. Gaze-based monitoring can be applied at either the shoulder surfer or the device owner side. At the shoulder surfer side, the system can decide whether an onlooker is deliberately looking at the screen by monitoring the onlooker's eye movements. For instance, Zhou et al. \cite{Zhou:EMC:2016} proposed a shoulder surfing detector using motion tracking sensors to locate and orient the onlookers close to the device. When the system detects that a bystander is gazing at the screen, it pops up multiple visual and auditory notifications to raise attention to the device owner. Regarding the device owners, gaze-based monitoring can be leveraged to track their eye movements to decide what content they are focusing on in real-time. For instance, Ragozin et al. \cite{Ragozin:PRU:2019} proposed a private digital reading approach employing an eye-tracker to decide the content being watched by the device owner. Only this portion of the content is visible, while others are obfuscated automatically and instantly.

Gaze-based monitoring is a promising direction for privacy-preserving image viewing at the end devices. However, there exist several challenges at present. The device owner-side solutions make only the contents the device owner staring at visible, which is less feasible for image data since users tend to enjoy an image in its entirety. The other set of solutions trying to track shoulder surfers' behaviors may bring its own privacy concerns, given that the monitoring is performed without the bystanders' permission.

\subsubsection{Solutions for third-party applications}
The third-party applications can be divided into two groups: the honest applications and the dishonest ones, upon which the privacy intelligence solutions differ. The honest applications are assumed to be willing to respect individual privacy by actively improving their backend model to be privacy-preserving. This improvement can be made by adversarial learning the backend model. Regarding the dishonest applications coveting the valuable private information of user photos, secure multi-party communication can be leveraged to block their malicious actions while maintaining legitimate social functionality. 

\paragraph{\textbf{Adversarial learning.}} Adversarial learning is a type of machine learning framework designed with adversarial goals: ensuring the availability of the desired recognition task for social functionality while incapacitating the models used for malicious recognition by attackers. For example, Ren et al. \cite{Ren:LAF:2018} proposed an adversarial learning algorithm for privacy-preserving life event monitoring. The algorithm involves two competing components: an anonymizer that modifies the original image to remove privacy-sensitive information while maintaining high-performance spatial action detection; and a discriminator that tries to extract privacy-sensitive information from the anonymized images. The competition between the two components results in a photo anonymizer that can anonymize human faces with barely affecting the accuracy of action detection. Wu et al. \cite{Wu:TPV:2018} provided a universal adversarial learning framework without the need to specify the recognition task of the third-party application. The motivation was that strong privacy protection should be sustainable against arbitrary attack models. The authors developed a learnable degradation transformation for the original image and proposed several training strategies. The resulting degradation makes the image resistant to unseen attack models while usable for legitimate social functionalities. 

\paragraph{\textbf{Secure multi-party computation}}
Secure multi-party computation aims to prevent client-side images from being directly exposed to the server in an untrusted environment. The client images are normally projected into a minimal feature set before being shared with the server to ensure legitimate social functionality while reducing unnecessary disclosure. 

\textit{Secure image recognition.}\indent Rahulamathavan et al. \cite{Rahulamathavan:EPF:2017} proposed a privacy-preserving algorithm for facial expression classification in a mutually-untrusted client-server scenario where neither the client nor the server wants to reveal the inputs and outputs to each other. The authors proposed a lightweight algorithm that projects the image onto a low-dimensional feature space in private with a randomization mechanism. Then, the expression is classified by feature distance matching on the anonymized features. In this way, the client-side images and the server-side classification results can be mutually anonymized. Nakamura et al. \cite{nakamura2018encryption} developed a general privacy-preserving client-server image recognition framework to avoid the server to know the recognition result. First, client users extract a visual feature from their taken photo and transform it so that the server cannot uniquely determine the recognition result. Then, the users send the transformed feature to the server that returns a set of candidates of the recognition result to the users. Finally, the users compare the candidates to the original visual feature for obtaining the final result. 

\textit{Secure image retrieval.}\indent Some studies focus on privacy-preserving cloud image retrieval using secure multi-party computation. For example, Xia et al. \cite{xia2020privacy} proposed a privacy-preserving image retrieval scheme, in which the images are encrypted but similar images to a query can be retrieved from the encrypted images. Secure Local Binary Pattern (LBP) features can be directly extracted as the local features from the encrypted images without server-client communication for similarity matching. Zhang et al. \cite{zhang2020privacy} proposed a privacy-preserving scheme for content-based image retrieval and sharing in cloud-based social multimedia applications. The users extract visual features from the images, and perform locality-sensitive hashing functions on visual features to generate image profile vectors. A secure index structure based on cuckoo hashing is then designed for profile vector matching.

\subsubsection{Solutions for image deletion}
\paragraph{\textbf{Digital oblivion}} 
Given the free dissemination of personal information and the availability of cheap and massive online digital storage means, OSN may "remember" the shared images even if the user has proactively deleted the images from the original social circle. Digital oblivion stands for a set of automated online image deletion techniques.

\textit{Self-destruction-based oblivion.}\indent Some digital oblivion solutions employ a data self-destruction mechanism by injecting an invisible deletion trigger into the image. For example, Backes et al. \cite{Backes:XDE:2011} developed an algorithm allowing users to embed an adjustable expiration date into OSN images. The modified images become inaccessible automatically once the expiration date has been reached, without users performing additional interaction with any PSP. Yang et al. \cite{Yang:PGP:2020} proposed selectively removing photos from a photo gallery to reduce inter-photo relevance on geolocation information. The goal was to decide the minimal set of images for removal from the collection to ensure that the true location was unpredictable in the remaining images. The authors formulated this collection censoring task as a combinatorial optimization problem and resolved it using the mixed-integer linear programming algorithm.  

\textit{Collaboration-based oblivion.}\indent Some studies have leveraged collaborative mechanisms for digital oblivion. Domingo-Ferrer et al. \cite{Ferrer:RED:2011} designed a set of protocols based on game theory to encourage users who receive information from an individual to rationally help the individual enforce his/her oblivion policy. Different fingerprints for different receivers are added to the content so that the owner can trace any unlawful use or spread of the image after the expiration date has passed. Stokes et al. \cite{Stokes:PAC:2013} designed a system enabling the peer-to-peer (P2P) agent community to assist in digital oblivion within OSNs. This is a P2P community made up of individuals who agree to protect the privacy of individuals who request that certain images be forgotten. The system involves a family of protocols to maintain up-to-date information on oblivion requests, and implements filtering functionality based on the authentication of user-to-content relations that are particularly relevant to digital oblivion. 

Currently, digital oblivion solutions for deletion delay typically require users to specify an expiration date as a deletion trigger and embed such information within the image file as implicit watermarks or fingerprints. The challenge is in managing the increasing volume of personal information shared and stored online. Users would benefit from more intelligent support for digital oblivion other than pre-defined rules, which would assure the long-term tracking of disclosed information and automatically safeguard users from information relating to a past episode surfacing unexpectedly \cite{Oblivion}. Future intelligent digital oblivion designs may be inspired by consensus-based mechanisms, e.g., a blockchain-based deletion scheme \cite{blockchain}, which leverages the blockchain technique to build a trusted P2P chain for data deletion.



\begin{table}[htbp]
\footnotesize
  \centering
  \caption{A summary of the solutions of privacy intelligence in the social experience stage. OP: observable privacy; IP: inferential privacy; CP: contextual privacy. PU: privacy-utility trade-off; PE: personalization; IN: independence; AU: automation; FL: flexibility; CE: communication-effectiveness. The ranking system is explained in Section \ref{ranking}.}
    \begin{tabular}{l|l|l|l|l|l|l|l|l|l|l}
    \toprule
    \multicolumn{1}{p{8em}|}{Privacy intelligence } & \multicolumn{1}{p{2em}|}{Paper; Year} & \multicolumn{1}{p{6em}|}{Sub-class} & \multicolumn{1}{p{3.5em}|}{Target privacy} & \multicolumn{1}{p{7em}|}{Key technique} & \multicolumn{1}{p{1.5em}|}{\textit{G1}-PU} & \multicolumn{1}{p{1.5em}|}{\textit{G2}-PE} & \multicolumn{1}{p{1.5em}|}{\textit{G3}-IN} & \multicolumn{1}{p{1.5em}|}{\textit{G4}-AU} & \multicolumn{1}{p{1.5em}|}{\textit{G5}-FL} & \multicolumn{1}{p{1.5em}}{\textit{G6}-CE} \\
    \midrule
    \multicolumn{1}{r|}{\multirow{2}[4]{8em}{Privacy-respecting browsing}} & \cite{Zezschwitz:YCW:2016}; 2016 & -   & OP    & Image processing & \Circle   & \textbf{N/A}   & \CIRCLE & \CIRCLE & \CIRCLE & \Circle \\
\cmidrule{2-11}          & \cite{Tajik:BIP:2019}; 2019 & -   & OP; IP; CP & \makecell[l]{Block encryption} & \CIRCLE & \textbf{N/A}   & \CIRCLE & \CIRCLE & \LEFTcircle   & \CIRCLE \\
    \midrule
    \multicolumn{1}{r|}{\multirow{2}[4]{8em}{Gaze-based monitoring}} & \cite{Zhou:EMC:2016}; 2016 & -  & OP    & Motion tracking & \textbf{N/A}   & \textbf{N/A}   & \Circle   & \CIRCLE & \CIRCLE & \Circle \\
\cmidrule{2-11}          & \cite{Ragozin:PRU:2019}; 2019 & -   & OP    & Eyeball tracking & \textbf{N/A}   & \textbf{N/A}   & \Circle   & \CIRCLE & \CIRCLE & \Circle \\
    \midrule
    \multicolumn{1}{r|}{\multirow{2}[4]{8em}{Adversarial learning}} & \cite{Ren:LAF:2018}; 2018 & -  & OP;   & \makecell[l]{Generative neural \\network} & \CIRCLE & \Circle   & \CIRCLE & \CIRCLE & \Circle   & \textbf{N/A} \\
\cmidrule{2-11}          & \cite{Wu:TPV:2018}; 2018 & -   & OP; IP & \makecell[l]{Generative neural \\network} & \CIRCLE & \Circle   & \CIRCLE & \CIRCLE & \Circle   & \textbf{N/A} \\
    \midrule
    \multicolumn{1}{l|}{\multirow{4}[8]{8em}{\makecell[l]{Secure multi-party \\ computation}}} & \cite{Rahulamathavan:EPF:2017}; 2017 & \makecell[l]{Secure image \\recognition} & IP    & Secure protocol & \CIRCLE & \textbf{N/A}   & \CIRCLE & \CIRCLE & \LEFTcircle   & \CIRCLE \\
\cmidrule{2-11}          & \cite{nakamura2018encryption}; 2018 & \makecell[l]{Secure image \\recognition} & OP, IP & Secure protocol & \CIRCLE & \textbf{N/A}   & \CIRCLE & \CIRCLE & \LEFTcircle   & \CIRCLE \\
\cmidrule{2-11}          & \cite{xia2020privacy}; 2020 & \makecell[l]{Secure image \\retrieval} & OP    & Secure protocol & \CIRCLE & \textbf{N/A}   & \CIRCLE & \CIRCLE & \LEFTcircle   & \CIRCLE \\
\cmidrule{2-11}          & \cite{zhang2020privacy}; 2020 & \makecell[l]{Secure image \\retrieval} & OP    & Secure protocol & \CIRCLE & \textbf{N/A}   & \CIRCLE & \CIRCLE & \LEFTcircle   & \CIRCLE \\
    \midrule
    \multicolumn{1}{r|}{\multirow{4}[8]{8em}{Digital oblivion}} & \cite{Backes:XDE:2011}; 2011 & \makecell[l]{Self-destruction\\-based oblivion} & OP; IP; CP & Rule design & \textbf{N/A}   & \CIRCLE & \CIRCLE & \CIRCLE & \CIRCLE & \textbf{N/A} \\
\cmidrule{2-11}          & \cite{Yang:PGP:2020}; 2020 & \makecell[l]{Self-destruction\\-based oblivion} & IP    & Machine learning & \Circle   & \textbf{N/A}   & \Circle   & \CIRCLE & \LEFTcircle   & \textbf{N/A} \\
\cmidrule{2-11}          & \cite{Ferrer:RED:2011}; 2011 & \makecell[l]{Collaboration\\-based oblivion} & OP; IP; CP & Rule design & \Circle   & \CIRCLE & \Circle   & \Circle   & \LEFTcircle   & \Circle \\
\cmidrule{2-11}          & \cite{Stokes:PAC:2013}; 2013 & \makecell[l]{Collaboration\\-based oblivion} & OP; IP; CP & Rule design & \Circle   & \CIRCLE & \Circle   & \Circle   & \LEFTcircle   & \Circle \\
    \bottomrule
    \end{tabular}%
  \label{tab:stage3}%
\end{table}%

\subsection{Design principles in social experience}
Table \ref{tab:stage3} provides a breakdown of the reviewed solutions in the social experience stage. Common design principles regarding OSN image privacy in this stage are as follows:

\begin{itemize}
	\item \textbf{In-the-wild mode.} In the social experience stage, images have been already accessed and are controlled by the recipients. In a sense, images are in the wild from now on, out of the original owner's control. Given the uncertainty of the recipients' behaviors, more open privacy issues are raised that is difficult for users to figure out one by one. OSN servers should play a more important role in protecting images from leakage. In this way, the reviewed solutions in this stage can be implemented as add-in plugs in the OSN services, allowing OSN servers to provide persistent preservation.
	
	\item \textbf{Contextual integrity.} The contextual integrity property of OSN image privacy should be a primary design goal in this stage, which confirms whether the information contained in the shared image is intact during social usage. The design goal can be interpreted from two angles according to whether the recipients are trusted or not. For a trusted recipient, the goal is to preserve privacy while ensuring the complete images' social usability for human viewers or functional availability for social applications. For untrusted recipients, the goal is to detect potential image manipulation and infringement to thwart information leaks.

    \item \textbf{Persistent intelligence.} As strong measures regarding privacy have been taken at the previous stage, the main target of privacy intelligence in this stage is to intensify image privacy further in the context of social experience to maintain a privacy-friendly environment in the long run. For this reason, we name privacy intelligence in this stage as \emph{persistent intelligence}. 
\end{itemize}



\section{Challenges and future direction}\label{section6}

\subsection{Modelling OSN image privacy}
Many intelligent solutions in this realm provide automated privacy prediction, management or recommendation mechanisms to assist users' decision-making process. These mechanisms mainly involve a privacy knowledge model, which is able to identify the statistical correlation between individual privacy and image data and OSN contextual factors. Currently, two significant factors are often underestimated in OSN image privacy modelling. One is the spatio-temporal factor which reflects the dynamic of privacy needs changing with time and location in different environments or situations \cite{privacy_regu}. Another factor is the incident factor, which indicates what is going on in the image, as different content has different degrees of sensitivity (even for the same participants). Incorporating the two factors in privacy pattern modelling requires a higher level of image understanding. The recent advance in deep learning-based social image understanding \cite{ImageUnderstanding1, ImageUnderstanding4} may be promising direction for this challenge. 

Although various privacy knowledge models have been proposed to satisfy different specific scenarios, we believe a universal modelling method will be more attractive in the future. To identify the prior knowledge required for a universal privacy model, we discover three boundaries of OSN image privacy inspired by the theory of privacy boundaries \cite{Palen:2003:UPN:642611.642635}, including: 

\begin{itemize}
    \item \textbf{The disclosure boundary}, which manages the tension between private and public, i.e., the degree of individual information disclosure from OSN image sharing in subjective self-cognition.

    \item \textbf{The identity boundary}, which manages the tension between self and other in the context of multi-party interactions. Individual privacy needs for OSN image sharing may vary depending on different representations of identity in different social groups. 

    \item \textbf{The spatio-temporal boundary}, which manages the tension of privacy decisions changing over time and location.
\end{itemize}


\subsection{Privacy-utility trade-off}
The privacy-utility trade-off is always an open challenge in designing privacy-enhancing techniques, especially in the OSN image sharing context where there is a specific purpose regarding the shared images. According to the literature, privacy and utility can be relatively well balanced in solutions relying on access control or encryption techniques since they make little modification directly to the image. However, this problem is still challenging in solutions requiring image processing, such as visual obfuscation or adversarial perturbation, since processing an image will damage its natural information integrity to some extent.

\begin{figure}[htbp]
\centering
\includegraphics[width=0.9\textwidth]{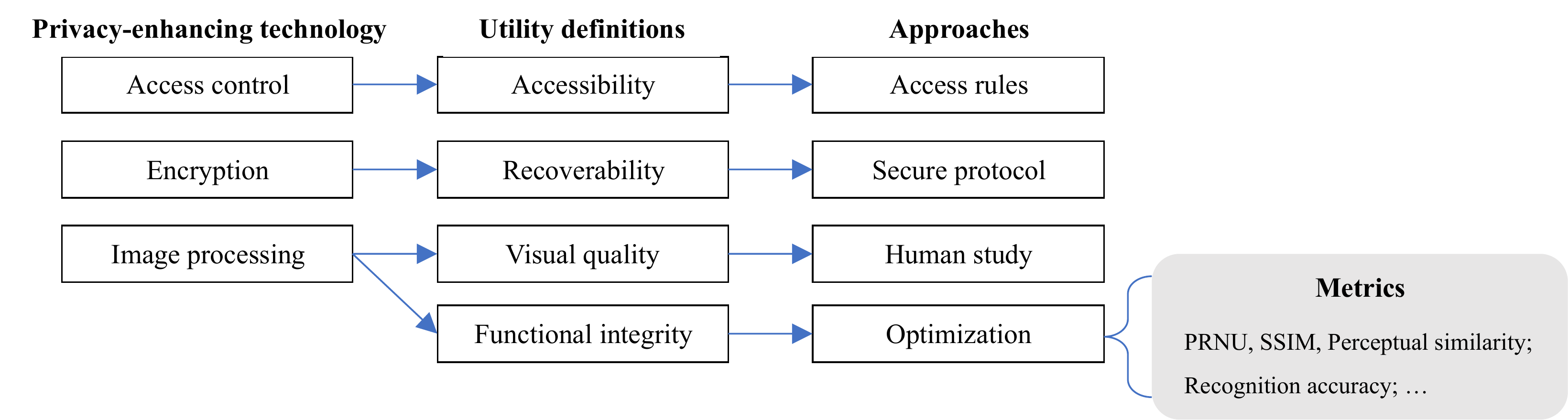}
\caption{The difference in treating image utility by different privacy-enhancing techniques.}
\label{utility}
\end{figure}

Ensuring the privacy-utility trade-off requires an appropriate formulation of the utility loss. Figure \ref{utility} shows the difference in treating image utility with various privacy-enhancing techniques. Image utility is normally defined as accessibility in access control techniques, and the trade-off is satisfied by designing specific access rules \cite{ding2019extended}. In encryption techniques, utility is often guaranteed by recoverability, where the authenticated users can recover the original image by lawful keys. Regarding image processing techniques, some studies defined the utility as visual quality and evaluated it via subjective human ranking scores \cite{Li:BBI:2017, Li:EUE:2017, Hasan:VEO:2018, Hasan:CPB:2019}. In contrast, more studies prefer to formulate utility as an optimization objective with specific quantifying metrics. For example, photo response non-uniformity (PRNU), structural similarity (SSIM), and perceptual loss \cite{PL:Justin} are common metrics for visual quality, and recognition accuracy is widely used for assessing functional integrity. In this sense, a promising direction for balancing the privacy-utility trade-off is to optimize the two objectives simultaneously via an adversarial game \cite{Raval:PVS:2017}. Moreover, existing measurements of utility loss are mostly evaluated on static images. The measurements in a dynamic OSN context requires more specific considerations on users' interests and behaviors, which remains further investigation.

\subsection{DeepFake: challenging the real world}
DeepFake is an emerging face forgery technique powered by deep learning, by which an attacker with little image processing knowledge can produce a realistic fake media record based on the victim’s face photos. DeepFake can effectively generate natural and realistic fake faces from a real face photo and seamlessly blend them into other media records. It is a image security issue rather than a privacy issue, but can be more severe along with private image leakage. 

Nowadays, there is widespread concern about the malicious application of DeepFake, due to its potentially disruptive consequences to visual security, laws, politics, and society in general \cite{deepfakeDam1}. The research community has become an influential force in motivating studies on DeepFake detection and anti-detection \cite{deepfakeSurvey1, deepfakeSurvey2, liu2022making}. Multiple large-scale DeepFake detection datasets have been released, such as FaceForensics++ \cite{DeepFakeDataset2} and Celeb-DF \cite{DeepFakeDataset3}. However, pursuing detection as the only solution may be insufficient, as this is a retroactive countermeasure implemented after the attacks have already had their effect. Moreover, it is extremely hard to defend against DeepFake attacks once malicious viewers have thoroughly accessed face photos because, by then, they are capable of fully controlling the data. Therefore, we believe more effort is needed in forestalling and preventing malicious users from getting the data. More investigations are needed on how to identify malicious users according to the historical internet trails before sharing images \cite{horng2011novel, hu2022privacy}. 


\subsection{The right to be forgotten}
Unlimited retention of personal images on the web may harm individual privacy. For example, teenagers may suffer long-term disadvantages to their future life and career due to indiscreet photos shared on social media. In the long run, many users desire to dissociate themselves from obsolete information that represents their past identity and behaviors. Therefore, the right to be forgotten, as a critical clause in the GDPR \cite{Victor:GDPR}, should be guaranteed in a privacy-friendly environment for OSN image sharing.

One considerable challenge to ensure the right to be forgotten is that the shared images are normally associated with multiple information sources. On the one hand, one photo of a user may be correlated with a collection of photos owned by other users, such that a simple deletion on the single user side cannot thoroughly erase the sensitive information. On the other hand, user data related to image content is easily exchanged across multiple ad hoc social networks. For example, one's private presence can be recorded and shared simultaneously by personal photography (shared in the OSN domain) and location information (shared in the vehicular social network domain \cite{horng2015efficient,tzeng2015enhancing}). Such cross-domain relations pose intractable challenges to achieving the right to be forgotten by only deleting image data from the OSN domain.

\subsection{Paradox of privacy dataset publishing}

Most ML-based intelligent solutions, such as learning-based privacy prediction and personalized policy generation, are essentially data-driven, heavily relying on datasets with image privacy knowledge. According to our literature review, only five image privacy datasets are publicly available, as shown in Table \ref{tab:dataset}. One cause leading to the scarcity of image privacy datasets is the paradox of privacy dataset publishing, which means that a privacy-relevant dataset naturally contains certain sensitive information and thus should not be fully released to the public. Although some image privacy datasets only release abstract features or masked images as a countermeasure, models will suffer from a certain performance compromise if trained with the incomplete datasets.

\begin{table*}[!htbp]
\scriptsize
  \centering
  \caption{Details of publicly available image privacy datasets.}
    \begin{tabular}{p{7em}p{6.915em}p{5.585em}p{12.75em}p{11em}p{11em}}
    \toprule
    \textbf{Dataset \& Year} & \textbf{Image source} & \textbf{Annotation level} & \textbf{Available annotations} & \textbf{Dataset size} & \textbf{Remark} \\
    \midrule
    \textit{PicAlert} \cite{picalert}, 2012 & Internet (Flickr) & Image level;\newline{}Text level & Privacy category (binary);\newline{}User tags & N = 32106 (4701 private, 27405 public) & 1. Some images are expired;\newline{}2. Limited data modality \\
     \midrule
    \textit{YourAlert} \cite{youralert}, 2016 & Local collection (from 27 social network users) & Image level; & Privacy category (binary) & N = 1511 (444 private, 1067 public) & Only image features are released \\
     \midrule
    \textit{VISPR} \cite{VISPR}, 2017 & Internet (Flickr and Twitter) & Object level & Privacy attribute (68 types) & N = 22167 (5.22 attributes per image) & Limited data modality \\
     \midrule
    \textit{VISPR-extension} \cite{vispre}, 2018 & Internet (Flickr and Twitter) & Object level;\newline{}Pixel level & Privacy attribute (24 types);\newline{}Attribute category (3 classes);\newline{}Private region; & N = 22167 (8473 images with region pixel-labeling) & \multicolumn{1}{l}{} \\
     \midrule
    \textit{VizWiz-Priv} \cite{VizWiz-Priv}, 2019 & Local collection (from blind photographers) & Image level;\newline{}Object level;\newline{}Pixel level;\newline{}Text level & Privacy attribute (23 types);\newline{}Private region;\newline{}Image/question pairs & N = 13630 (5537 private, 8093 publc); \newline{}5537 images with region pixel-labeling;\newline{}2685 image/question pairs & 1. Collected from a special group;\newline{}2. Only masked images are released \\
    \bottomrule
    \end{tabular}%
  \label{tab:dataset}%
\end{table*}%

An intuitive solution for disentangling the paradox  is to purchase the right to use from the private image owners. Then, privacy would be valued as a commodity and price would become the most important factor in a buyer-seller game. The privacy pricing problem can be motivated by some previous studies \cite{sellData1, sellData2, sellData3, sellData4}, which provides an auction-based trading mechanisms for private data. Another solution is to use alternative learning fashions such as distributed learning  \cite{distributedlearning,li2021bpt} and unsupervised or semi-supervised learning \cite{unsupervisedLearning1, unsupervisedLearning2, unsupervisedLearning3}, which means the raw data does not need to be accessed.

\section{Conclusion}\label{section7}
With a focus on the urgent privacy needs in modern OSN image sharing, we conducted a survey on privacy intelligence in such a sharing context, which is a collective term referring to intelligent solutions to various modern privacy issues derived from sharing-related user behaviors. Specifically, we first introduced the definition and taxonomy of OSN image privacy under the contextual constraints of dynamic OSN image sharing. Then, to analyze multiple privacy issues, solutions and challenges in this interdisciplinary area comprehensively, we proposed a high-level privacy analysis framework based on the entire lifecycle of OSN image sharing. Using the framework, we systematically identified modern privacy issues induced by OSN users' behaviors and explored the corresponding intelligent solutions in a stage-based fashion. For the reviewed intelligent solutions at each stage, we elaborated on their methods, advantages and disadvantages, and summarized their common design principles. We also discussed the challenges and future directions in this field. 

The privacy intelligence solutions explored in this survey are sufficient to form an intelligent privacy firewall that may contribute to building a more intelligent environment for privacy-friendly OSN image sharing with respecting the privacy of all stakeholders. We hope our work can facilitate current-day privacy management and reconcile the ever-increasing use of OSN image sharing and the modern individual privacy needs.


\begin{acks}
This paper is supported by an ARC project, LP180101150, from the Australian Research Council, Australia.
\end{acks}
\bibliographystyle{unsrt}
\bibliography{reference}

\end{document}